\documentclass[preprint]{imsart}
\RequirePackage[OT1]{fontenc}
\RequirePackage{amsthm,amsmath}
\RequirePackage[numbers, authoryear]{natbib}
\usepackage[colorlinks=true, linkcolor=blue, urlcolor=blue, citecolor=blue]{hyperref}
\usepackage{booktabs}           
\usepackage{graphics}          
\usepackage{graphicx}          
\usepackage{enumitem}
\startlocaldefs
\numberwithin{equation}{section}
\theoremstyle{plain}
\endlocaldefs

\begin{document}

\begin{frontmatter}

\title{Evaluating NHL Goalies, Skaters, and Teams Using Weighted Shots}
\runtitle{Weighted Shots}

\begin{aug}
\author{\fnms{Brian} \snm{Macdonald}\thanksref{t1} 
\ead[label=e1]{bmac@jhu.edu}} 
\author{\fnms{Craig} \snm{Lennon}\thanksref{t2} 
\ead[label=e2]{}}
\author{\fnms{Rodney} \snm{Sturdivant}\thanksref{t1}
}

\thankstext{t1}{Department of Mathematical Sciences, United States Military Academy, West Point, NY}
\thankstext{t2}{Army Research Lab, Aberdeen, MD}
\runauthor{B. Macdonald et al.}

\affiliation{Department of Mathematical Sciences, United States Military Academy\thanksmark{m1} and Army Research Lab\thanksmark{m2}}


\end{aug}

\begin{abstract}
In this paper, we develop a logistic regression model to estimate the probability that a particular shot in an NHL game will result in a goal, and use the results to evaluate the performance of NHL skaters, goalies, and teams.  We weight each shot based on the estimated probabilities obtained from our model, call this statistic ``weighted shots'', and use advanced statistics based on weighted shots as the basis of our evaluation.  We also analyze whether advanced statistics based on weighted shots outperform traditional statistics as an indicator of future performance of skaters, goalies, and teams.  In general, statistics based on weighted shots perform well, but not better than traditional statistics.  
We conclude that weighted shots should not be viewed as a replacement for those statistics, but can be used in conjunction with those statistics.  Finally, we use weighted shots as the dependent variable in an adjusted plus-minus model.  The results are estimates of each player's offensive and defensive contribution to his team's weighted shots during even strength, power play, and short handed situations, independent of the strength of his teammates, the strength of his opponents, and the zone in which his shifts begin.  
\end{abstract}

\begin{keyword}[class=AMS]
\kwd[Primary ]{62P99}
\kwd[; secondary ]{62J12}
\end{keyword}


\begin{keyword}
\kwd{logistic regression}
\kwd{hockey}
\kwd{sports}
\kwd{weighted shots}
\end{keyword}

\end{frontmatter}

\section{Introduction}
        The purpose of a shot quality model is to estimate the probability that a shot will be a goal based on detailed information about that shot.  These details can include distance from the goal, shot type (slap shot, wrist shot, backhand, etc.), game situation (power play, even strength, short handed), whether or not a shot was a rebound shot (meaning, the shot came very soon after another shot), or even the $(x,y)$ coordinate on the ice where the shot was taken.

        Once the probability that a shot will be a goal is estimated, the shot can be weighted based on that probability.  These ``weighted shots'', as we will call them, can be interpreted as the expected number of goals that will result from each shot.  
        Weighted shots have many uses, and here are three examples:
        
            \begin{enumerate}[leftmargin=.5cm]
                \item A team's defense can be rated based on the quality of shots the team allows by using weighted shots allowed.  Such a rating would be independent of the strength of the team's goalies.  
                \item A goalie can be rated based on ``adjusted save percentage''.  This statistic is the difference between his actual save percentage and his ``expected save percentage'', which is based on the quality of shots that he faces.  Such a rating would be independent of the team's defense.
                \item Weighted shots per 60 minutes can be used as the outcome variable in an adjusted plus-minus model instead of goals per 60 minutes.  The results from such an adjusted plus-minus model would be estimates of a player's contribution to his team in terms of weighted shots per 60 minutes, independent of the strength of the player's teammates and opponents, and independent of the zone in which his shifts begin.  
            \end{enumerate}
   
        In this paper, we describe a new weighted shots model and use the results to analyze the performance of skaters, goalies, and teams.  In Section $\ref{models}$, we briefly summarize the weighted shots models of \cite{ken1}, \cite{ken2}, \cite{ken3}, \cite{awadwshot}, and \cite{digr}, including defining many of the variables they used.  In Section $\ref{newvariables}$ we describe some new variables quantifying the fatigue of the shooter as well as the average time that the offense and defense have been on the ice at the time of the shot.  We also describe the interaction terms we will use in our model in that section.  In Section $\ref{logistic}$, we describe the results of our logistic regression model.  The area under the Receiver-Operator Characteristic (ROC) curve is 0.764, which indicates that the model performs well at predicting that a shot will be a goal, and justifies the use of an ``adjusted save percentage'' statistic to describe the past performance of goalies.  Then, we describe how weighted shots, and statistics based on weighted shots, can be used to evaluate skaters, goalies, and teams in Section $\ref{teamperformance}$.  Since some of these applications were already thoroughly covered by \cite{johns},  \cite{ryder-shot-quality}, \cite{ken1}, \cite{ken2}, and \cite{ken3}, \cite{awadwshot}, or \cite{digr}, we focus on new analysis.  In particular, we analyze the reliability and predictive power of some advanced statistics based on weighted shots, and we find the following: 
            \begin{enumerate}[leftmargin=.5cm]
                \item At the goalie level, we find \textit{some} evidence that adjusted save percentage is a better measure of performance than save percentage, is more consistent than save percentage, is a better predictor of future performance than save percentage, and is a better predictor of future save percentage than save percentage.  However, the evidence is far from overwhelming, and firm conclusions should not be drawn from this analysis.  Also, any potential improvement in predictive performance gained by using adjusted save percentage is small. (See Section \ref{section-adj-save}.)
                \item At the team level, adjusted shooting percentage does not seem to perform any better than (unadjusted) shooting percentage. (See Section \ref{section-team-stats}.)
                \item We did not find evidence that weighted shots per 60 minutes generated by a team's offense or allowed by a team's defense performed better than shots, Fenwick (shots + missed shots), or Corsi  (shots + missed shots + blocked shots) per 60 minutes.  However, weighted shots still performed well, and it is reasonable to use weighted shots per 60 minutes in conjunction with (but certainly not as a replacement for) those statistics. (See Section \ref{section-team-stats}.)
        \end{enumerate} 
        In Section \ref{apmw}, we use weighted shots in an adjusted plus-minus model to evaluate each NHL player's contribution to his team, independent of the strength of his teammates, the strength of his opponents, and the zone in which his shifts begin.  Finally, we finish with some ideas for future work and some conclusions in Section $\ref{conclusions}$.

\section{Brief Summary of Existing Models}\label{models}

    We start by defining some of the explanatory variables that we will use in our model which were also previously used by Krzywicki, Awad, or Schuckers, or used by all three.  Data for these variables can be found in the play-by-play files on websites like NHL.com or ESPN.com. 
   
        \begin{description}[leftmargin=.2cm]
          \item[Distance] The distance of the shot from the goal.  (See Figure $\ref{angle-image}$.)
              \begin{figure}[h!]
              \centering
              \includegraphics[width=.45\linewidth]{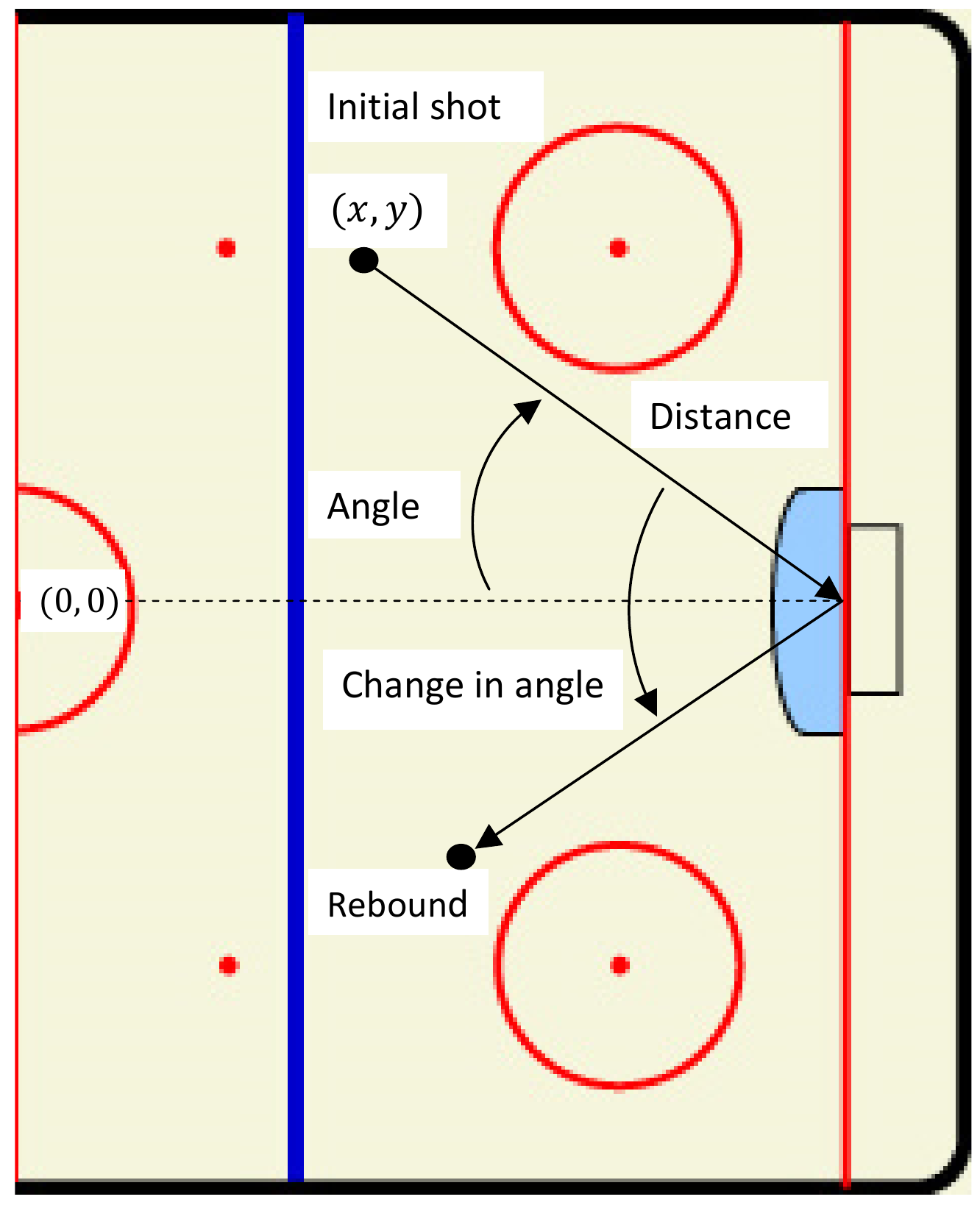}
              \caption{A visual explanation of distance, shot angle, and change in angle for rebound shots.}
              \label{angle-image}
              \end{figure}
          
          \item[Shot angle] The angle at which the shot was taken.  Directly in front of the goalie, in the middle of the ice, is an angle of $0$.  One would expect that shots taken with an angle near $0$ would be more likely to go in.  Near the side boards, on the goalie's left or right, is a large positive angle, and one would expect that shots from those locations would be less likely to be a goal.  The information about angle is obtained from the $(x,y)$ coordinates that describe the location of each shot on the ice, along with the dimensions of an NHL rink.  See Figure $\ref{angle-image}$ for a visual depiction of shot angle. 
          \item[Rebound] A shot within two seconds of another shot, with a distance less than $25$ feet, and no intervening event  
          \item[Own rebound] Rebound shot taken by the player that took the initial shot
          \item[Situation] Even strength, power play, or short handed
          \item[Shot type] Wrist shot, wrap-around shot, slap shot, backhand shot, snap shot, tip-in.  The reference category is wrap-around shot.
          \item[Score] Shooting team's score minus the defending team's score
          \item[Home Team] Indicates if a shot was taken by the home team
        \end{description}

        We note that instead of using shot angle, Schuckers uses $(x,y)$ 
        coordinates directly, and uses a non-parametric regression model instead of a logistic regression model.  Krzywicki and Awad use a logistic regression model, which is what we use here.  In his latest paper, Krzywicki had the creative idea to include a ``push'' variable, which we will call ``Angle Change'':
        \begin{description}[leftmargin=.2cm]
        \item[Angle Change]  For a rebound shot, Angle Change measures how far the goalie has to move from left to right or right to left to save the rebound shot.  Angle Change is measured in degrees, and can be thought of as the change in angle between the initial shot and the rebound shot.  For example, if the initial shot is $45$ degrees to the right of the goalie, and the rebound shot is $45$ degrees to the left, then Angle Change is $90$ degrees.  See Figure $\ref{angle-image}$.
        \item[Angle Change Left] Change in angle from right to left
        \item[Angle Change Right] Change in angle from left to right
        \end{description}

\section{New variables and interaction terms}\label{newvariables}
        We use the variables above along with the following additional variables:
        \begin{description}[leftmargin=.2cm]
          \item[Shooter fatigue] The length of time that the shooter has been on the ice during his current shift.  Data for this variable is obtained from the detailed shift reports on NHL.com.  
          \item[Shooting team time on ice] The average length of time, in seconds, that the shooting team's players had been on the ice at the time when the shot was taken.  For example, if all three of the shooting team's forwards were $10$ seconds into their shift when the shot was taken, and the shooting team's two defensemen had been on the ice for $15$ seconds when the shot was taken, then this variable would be the average of those times:  $(10+10+10+15+15)/5 = 60/5 = 12.$  
          \item[Defending team time on ice] Same as above, except for the team that did not take the shot.
          \item[Detailed Strength] Instead of even strength (EV), power play (PP), or short handed (SH),  we use more precisely classified situations:  
            EV $5$-on-$5$
          , EV $4$-on-$4$
          , EV $3$-on-$3$
          , PP $5$-on-$4$
          , PP $5$-on-$3$%
          , PP $4$-on-$3$%
          , SH $4$-on-$5$%
          , SH $3$-on-$5$%
          , and SH $3$-on-$4$.  There is too little data for $3$-on-$3$, so these situations were grouped with $4$-on-$4$.  Also, we found little difference between PP $4$-on-$3$ and PP $5$-on-$4$, so PP $4$-on-$3$ was included with PP $5$-on-$4$ as a ``one-man advantage'' variable.  PP $5$-on-$3$, however, remained separate.  For the same reasons, SH $3$-on-$4$ was grouped with SH $4$-on-$5$, and SH $3$-on-$5$ remained separate.  The reference category is EV 5-on-5.  
          
        \end{description}

    We included three interaction terms in the model, each dealing with angle: Angle on Rebounds, Angle on Own Rebounds, and Angle on Tip shots.  Typically, on shots from a large angle, the goalie can ``square up'' on a shooter, and the shooter has very little space in which to score.  We thought large angles would not be as bad for rebounds and tips because it is often more difficult for a goalie to be in good position for these types of shots.  Each of these interaction terms did turn out to be significant and remained in our final model.
     
\section{Logistic Regression Model}\label{logistic}
    We form a logistic regression model, which estimates the probably that a certain type of shot will result in a goal, using the variables mentioned above.  We use every shot during the 2008-09, 2009-10, and 2010-11 seasons that originated from the offensive zone, and for which the goalie was on the ice.  In Table $\ref{summary}$, we summarize the results of the logistic regression.  In the column labeled ``Coeff'', we give the estimated coefficients of the model.  We can tell how a variable affects the probability that the shot will be a goal by looking at the sign of the corresponding coefficient.  For example, the coefficient for PP$54$ is positive, which indicates that a shot on a $5$-on-$4$ power play is more likely to be a goal than a shot taken at even strength $5$-on-$5$.  The column ``Odds'' contains the odds ratio associated with each variable.  For example, the odds of a shot taken during a $5$-on-$4$ power play becoming a goal are $1.44$ times greater than for a shot taken at even strength $5$-on-$5$.  A odds of a shot during a $5$-on-$3$ power play becoming a goal is $2.53$ times greater.  
    
    Our new variable ``shooter fatigue'' is significant and has the sign we would expect.  The longer a shooter is on the ice before his shot, the more tired he is, and less likely his shot is to be a goal.  Interestingly, ``Defending Team Time on Ice'' is near zero and is marginally significant.  We would have expected that the longer the defending players are on the ice, the less likely they are to contest a shot, and the more likely that a shot would be a goal.  Also, ``Shooting Team Time on Ice'' is positive.  In these cases, there are possible correlations with power play variables and other variables that may explain the surprising signs of these coefficients.  For example, a shooting team on a power play that is on the ice for a minute may be a little tired, but they probably have the puck in the offense zone and are putting pressure on the other team's defense and goalie.  This variable may be quantifying offensive pressure rather than fatigue in a case like that.  

\begin{figure}[h!]
        \centering
        \includegraphics[width=.65\linewidth]{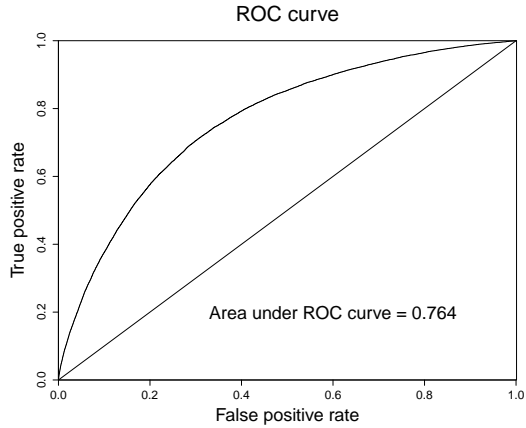}
        \caption{The ROC curve for our model}
                    \label{ROC-curve}
        \end{figure}

    The Receiver-Operator characteristic (ROC) curve for this model is given in Figure $\ref{ROC-curve}$.  
    The ROC curve is useful as a measure of the model's ability to discriminate between shots that were goals and those that were not.  An area under the ROC curve close to one means outstanding discrimination.  A value of 0.5, which corresponds to the diagonal line in Figure~\ref{ROC-curve}, indicates the performance is no better than a coin flip.  A value over 0.7 is generally considered good discrimination meaning the model is useful for determining the response for a given shot (see \cite{hosmer-lemeshow}, pages 160-163). The area under the ROC curve for our model is 0.764.

        \paragraph{Top 5 shots most likely to be a goal}
        In Table $\ref{topshots}$ we give the $5$ shots that our model determined were the most likely to become a goal.  In that table, we give some of the details of each shot, the estimated probability that the shot will be a goal according to our model (the column labeled ``P(Goal)''), the standard error in the estimated probability (Err), and the actual outcome of that shot (Event). Note that the columns Change, Left, Right correspond to the variables Angle Change, Angle Change Left, Angle Change Right, respectively.

  Notice that these shots were all taken within $13$ feet of the goal, and were all rebound shots.  Shot $\#1$ was the only shot taken at even strength on the list, but that shot was a slap shot taken from an angle of $0$ (directly in front of the goal), $13$ feet out, and had a high change in angle.  Also, the shooter had only been on the ice for 6 seconds, while the defending team had been on the ice for average of 44 seconds, and may have been fatigued and less able to contest the shot.  
  \begin{table}[h!]
  \begin{center}
  \caption{The top 5 shots most likely to be a goal.}
  \label{topshots}
  {\footnotesize
  \begin{tabular}{rrrrrrrrrrrr}
    \addlinespace[.3em] \toprule 
Rank   & Dist & Angle & Reb & Change & Left & Right & Strength & Type & P(Goal) & Err & Event \\ 
    \midrule 
  1 & 13 & 0 & 1 & 48 & 0 & 48 & EV55 & slap & 0.90 & 0.01 & SHOT \\ 
    2 & 11 & 63 & 1 & 126 & 0 & 126 & PP53 & tip & 0.90 & 0.02 & GOAL \\ 
    3 & 12 & 0 & 1 & 51 & 51 & 0 & PP53 & wrist & 0.90 & 0.01 & GOAL \\ 
    4 & 12 & 38 & 1 & 107 & 0 & 107 & PP53 & wrist & 0.90 & 0.01 & GOAL \\ 
    5 & 13 & 14 & 1 & 58 & 0 & 58 & PP53 & wrist & 0.89 & 0.01 & SHOT \\ 
     \bottomrule 
  \end{tabular}
  }
  \end{center}
  \end{table}
   The rest of the shots in the list were taken on a $5$-on-$3$ power play, and had a low angle, a high change in angle, or both.     
   These general trends continue if we look at the top $50$ shots as well.  The results are intuitive, as we would expect shots of these types should have a high probability of being a goal. 
            
        \section{Evaluating Skaters, Goalies, and Teams}\label{teamperformance}
            In the previous section we discussed how the model performs at the level of a single shot, but what we would really like to know is how weighted shots can be used by NHL front offices, analysts, and fans to analyze the perfomance of skaters, goalies, and teams. Some of this analysis has been covered by \cite{johns}, \cite{ryder-shot-quality}, \cite{ken1}, \cite{ken2}, \cite{ken3},  \cite{awadwshot}, or \cite{digr}.  For example, Krzywicki used the results of his logistic regression model to analyze the performance of goalies while adjusting for the quality of their team's defense.  Schuckers used a non-parametric regression model to estimate the probability of a goal and used that to rate goalies.  He coined his metric as Defense Independent Goaltender Rating (DIGR), which, as the name suggests, gives a rating for each goalie that is independent of the quality of shots allowed by his defense.  Awad used his shot quality model to evaluate the performance of skaters, using expected goals based on shot quality instead of goals in a plus-minus type statistic.
            
            For completeness, we repeat some of this analysis here.  However, we also focus on analysis that has not been done before, namely an analysis of the reliability and predictive power of various advanced statistics based on weighted shots.  In all cases, we consider only even strength 5-on-5 situations. 
            
        \subsection{Skater Performance}
            First, we use weighted shots and advanced statistics based on weighted shots to quantify two components of goal scoring ability of players: generating a high quantity of high quality of shots, and capitalizing on those shots.  In Table $\ref{shooters-goals}$, we list the top 5 players with the greatest difference between actual goals scored and expected goals scored based on their weighted shots during the 2008-09, 2009-10, and 2010-11 seasons.
            These players could be considered the players that were the ``best shooters'' during the last three seasons, or the players who best capitalized on the quality of shots they had.  Note that the columns denote expected goals (EG), actual goals (G), the difference between G and EG (DiffG), the standard error in DiffG (GErr), shots on goal (Shots), shooting percentage (Sh\% $=$ G/Shots), expected shooting percentage (ExpSh\% $=$ EG/Shots), adjusted shooting percentage (AdjSh\% $=$ league average shooting percentage plus DiffSh\%, the difference in Sh\% and ExpSh\%), and the standard error in AdjSh\% (SErr), respectively.  
\begin{table}[h!]
\begin{center}
\caption{The top 5 shooters at 5-on-5 during the 2008-09, 2009-10, 2010-11 seasons}
\label{shooters-goals}
{\footnotesize
\begin{tabular}{lrrrrrrrrrrrr}
  \addlinespace[.3em] \toprule 
Player & Pos & Team & EG & G & DiffG & GErr & Shots & Sh\% & ExpSh\% &  AdjSh\% & SErr \\ 
  \midrule 
Ilya Kovalchuk & LW & N.J & 39 & 67 & 28 & 0.1 & 478 & 0.140 & 0.082 &  0.138 & 0.0001 \\ 
  Alexander Semin & RW & WSH & 39 & 67 & 28 & 0.1 & 495 & 0.135 & 0.079 &  0.136 & 0.0001 \\ 
  Sidney Crosby & C & PIT & 45 & 70 & 25 & 0.1 & 435 & 0.161 & 0.104 &  0.137 & 0.0002 \\ 
  Jarome Iginla & RW & CGY & 44 & 64 & 20 & 0.1 & 577 & 0.111 & 0.076 &  0.115 & 0.0001 \\ 
  Steven Stamkos & C & T.B & 41 & 59 & 18 & 0.1 & 439 & 0.134 & 0.094 &  0.120 & 0.0002 \\ 
   \bottomrule 
\end{tabular}
}
\end{center}
\end{table}

        The DiffG column, which is actual goals minus expected goals, could be interpreted as the number of goals the player scored above what was expected based on the quality of his shots.  In other words, it could be considered the number of goals that could be attributed to the player's shooting ability.  Most observers would agree that these players all have elite shooting ability or finishing ability, and this statistic supports that belief.  We note that Alex Ovechkin, who did not make this top 5 list, was tied for sixth with 17. 
        
        We stress that this list is not an attempt to identify the best goal scorers in the league, but rather the best shooters.  Alexander Semin, for example, has one of the most dangerous shots in the league and his inclusion on this list supports that belief.  However, Semin is not typically among the top 5 goal scorers in the league. Shooting ability is just one component of goal scoring.
        
        Another component of goal scoring is generating a high quantity of high quality shots on goal.  In Table \ref{shooters-wshot}, we give the top 5 players in terms of expected goals scored based on their weighted shots.          
\begin{table}[h!]
\begin{center}
\caption{The top 5 skaters in weighted shots at 5-on-5 during 2008-09, 2009-10, and 2010-11}
\label{shooters-wshot}
{\footnotesize
\begin{tabular}{lrrrrrrrrrrrr}
  \addlinespace[.3em] \toprule 
Player & Pos & Team & EG & G & DiffG & GErr & Shots & Sh\% & ExpSh\% & AdjSh\% & SErr \\ 
  \midrule 
Corey Perry & RW & ANA & 65 & 63 & $-$2 & 0.1 & 594 & 0.106 & 0.109 &  0.077 & 0.0002 \\ 
  Alex Ovechkin & LW & WSH & 57 & 74 & 17 & 0.1 & 766 & 0.097 & 0.075 &  0.101 & 0.0001 \\ 
  Jeff Carter & C & PHI & 56 & 65 & 9 & 0.1 & 630 & 0.103 & 0.088 &  0.095 & 0.0001 \\ 
  Joe Pavelski & C & S.J & 54 & 41 & $-$13 & 0.1 & 519 & 0.079 & 0.103 &  0.055 & 0.0002 \\ 
  Eric Staal & C & CAR & 51 & 48 & $-$3 & 0.1 & 592 & 0.081 & 0.086 &  0.075 & 0.0001 \\ 
   \bottomrule 
\end{tabular}
}
\end{center}
\end{table}
Note that weighted shots contain information about both the quantity and quality of shots.  This list can then be interpreted as the players that took the best combination of quantity and quality of shots, regardless or whether or not they capitalized on their chances.  If instead we desired to find the players with the highest quantity of shots, we could simply use (unweighted) shots, while to find the players who took the highest quality of shots, we could use expected shooting percentage.  Note that Alex Ovechkin led the league in goals scored in 5-on-5 situations during the 2008-09, 2009-10, and 2010-11 seasons, so it is not surprising to see him among the league leaders in generating quality scoring chances (2nd in EG) and capitalizing on those chances (6th in DiffG).
        
\begin{figure}[h!] 
 \includegraphics[width=.45\linewidth]{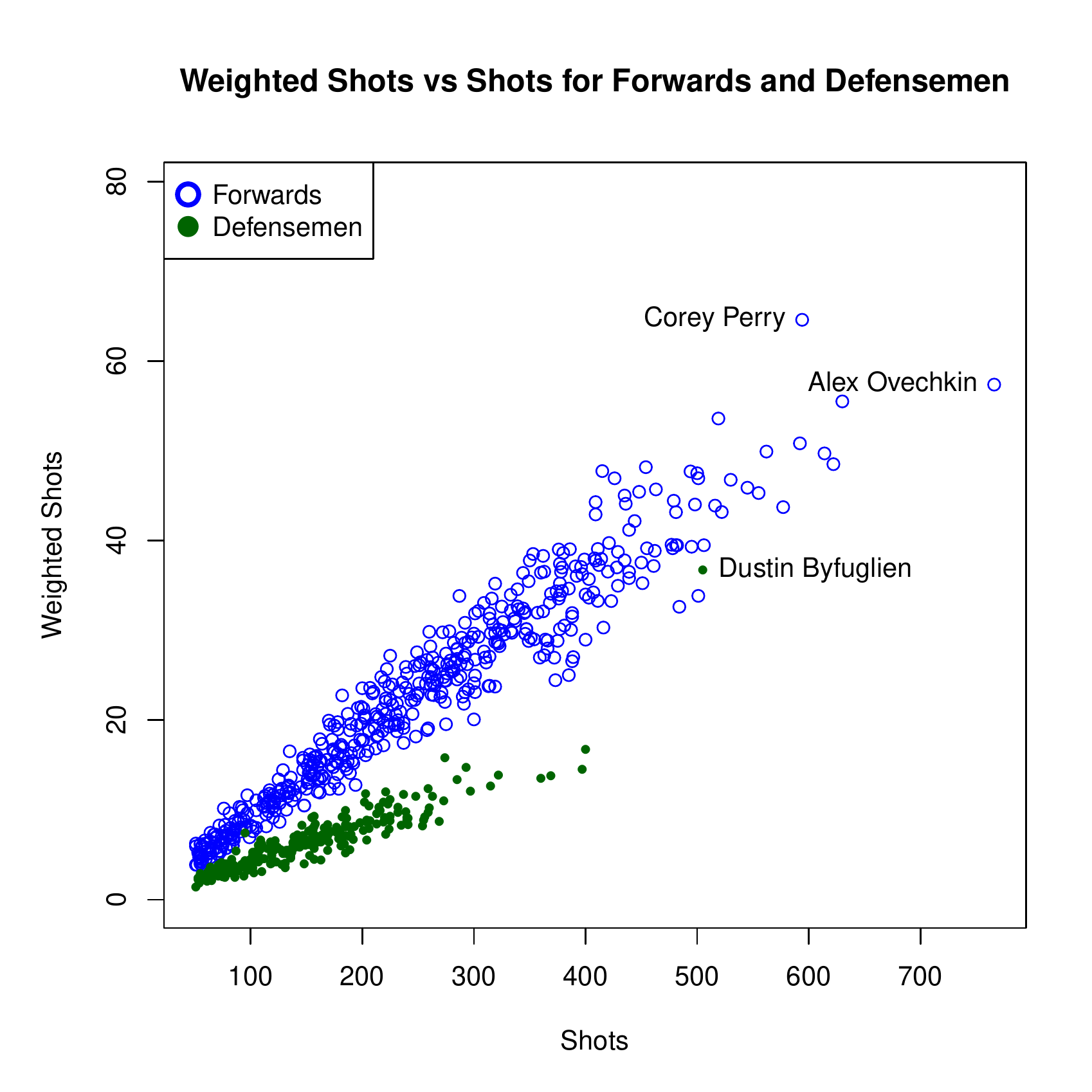}
 \includegraphics[width=.45\linewidth]{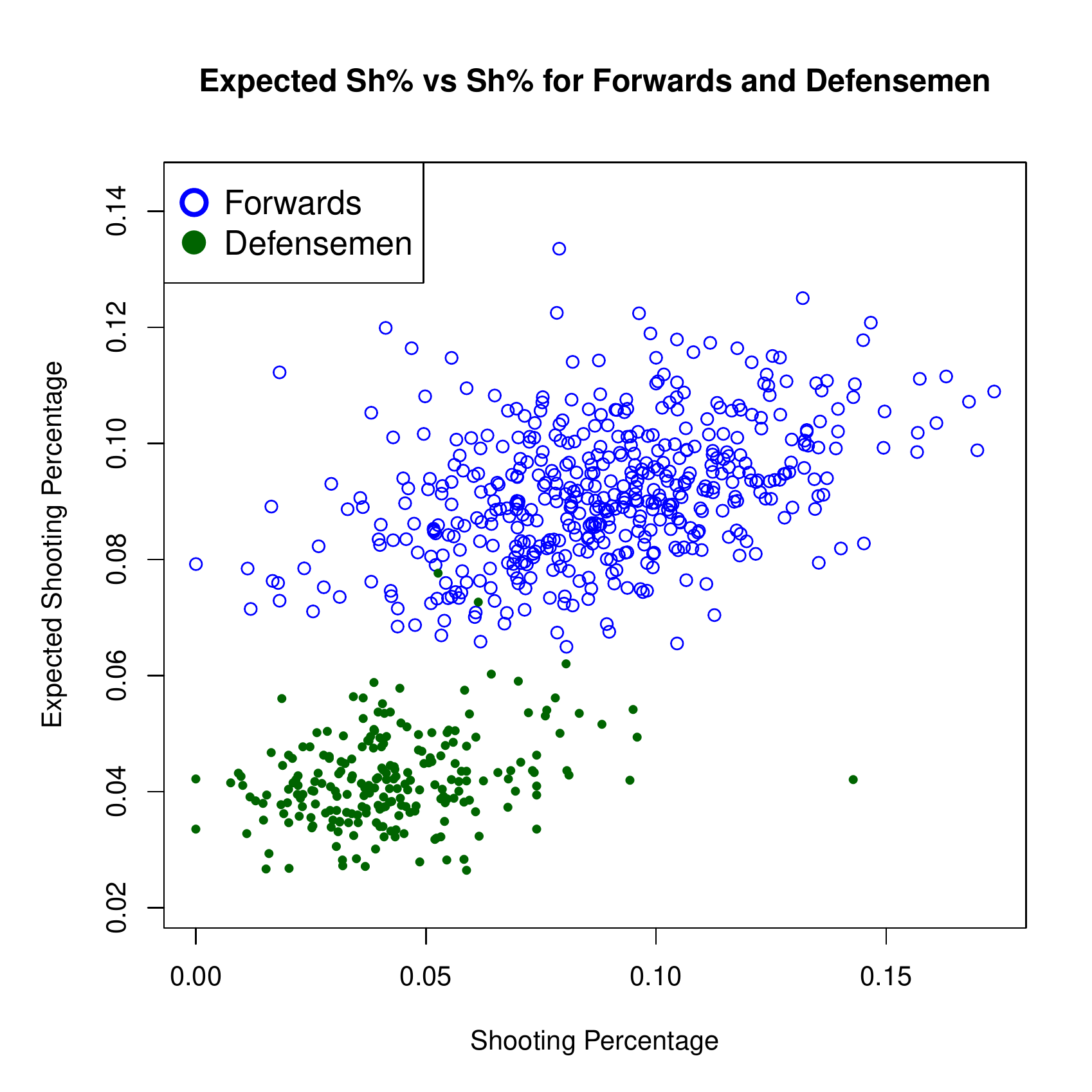}  
 \caption[]{(Left) Weighted shots versus shots for forwards and defensemen. (Right) Expected shooting percentage versus shooting percentage for forwards and defensemen (minimum 50 shots).}
 \label{figures/EV55-both-wshot-shots-F-and-D.pdf} 
  \end{figure}

        Since a big part of weighted shots is distance, the relationship between shots and weighted shots is much different for forwards than for defensemen. In Figure \ref{figures/EV55-both-wshot-shots-F-and-D.pdf}, we plot weighted shots versus shots for forwards (circles) and defensemen (dots).  
        There are two distinct groupings for forwards and defensemen, and shots by forwards are clearly weighted higher on average than shots by defensemen.  The single ``defenseman'' amongst the sea of forwards is Dustin Byfuglien, who was actually a forward during much of the 2008-09 and 2009-10 seasons when he was on the Chicago Blackhawks, and only started playing defense full-time after becoming an Atlanta Thrasher in 2010-11.  If we plot shooting percentage versus expected shooting percentage, as we do in the right half of Figure~\ref{figures/EV55-both-wshot-shots-F-and-D.pdf}, we get two groupings as well.  
        
        These observations suggest that forwards and defensemen should be treated separately if we desire to compare the reliability or predictive performance of weighted shots per 60 minutes with shots per 60 minutes.  
        We could find the correlation between weighted shots (respectively, shots) per 60 minutes in one half of a season and weighted shots (respectively, shots) per 60 minutes in the other half of a season as a way to measure how consistent the statistic is from one half of a season to another.  Additionally, we could find the correlation between weighted shots (respectively, shots) per 60 minutes in one half of a season and goals per 60 minutes in the other half of a season, as a way to quantify the ability of weighted shots (respectively, shots) to predict the goal scoring ability of a player.  We computed these correlations for forwards and defensemen separately and we found no evidence that weighted shots outperforms shots.  
        
\subsection{Goalie Performance}\label{goalieperformance}
    For goalies, we could think of weighted shots against as expected goals against, or the number of goals one would expect a goalie to allow given the quality of shots that he faced.  The difference between expected goals against and actual goals against could be interpreted as the number of goals that a goalie saved above what is expected.  In other words, it is a measure of a goalie's value in terms of goals prevented.  In Table $\ref{goalies-goals}$, we give the top 5 goalies in this measure from the $2010$-$2011$ season.  
\begin{table}[h!]
\begin{center}
\caption{The top 5 goalies in goals prevented in $2010$-$2011$.}
\label{goalies-goals}
{\footnotesize
\begin{tabular}{lrrrrrrrrrr}
  \addlinespace[.3em] \toprule 
Goalie & ExpGA & GA & DiffGA & GErr & ShotA & Sv\% & ExpSv\% & AdjSv\% & Err \\ 
  \midrule 
Tim Thomas & 145 & 103 & 42 & 0.2 & 1712 & 0.940 & 0.915 &  0.938 & 0.0001 \\ 
  Henrik Lundqvist & 179 & 148 & 31 & 0.2 & 1888 & 0.922 & 0.905 &  0.930 & 0.0001 \\ 
  Pekka Rinne & 152 & 125 & 27 & 0.2 & 1831 & 0.932 & 0.917 &  0.928 & 0.0001 \\ 
  Roberto Luongo & 145 & 120 & 25 & 0.1 & 1675 & 0.928 & 0.914 &  0.928 & 0.0001 \\ 
  Jonas Hiller & 131 & 108 & 23 & 0.2 & 1420 & 0.924 & 0.908 &  0.930 & 0.0001 \\ 
   \bottomrule 
\end{tabular}
}
\end{center}
\end{table}
We note that the columns correspond to expected goals against (ExpGA), actual goals against (GA), the difference in ExpGA and GA (DiffGA), the error in DiffGA (GErr), shots against (ShotA), save percentage (Sv\%  $=1 - \frac{GA}{ShotA})$, expected save percentage (ExpSv\% $=1-\frac{EGA}{ShotA})$, adjusted save percentage (AdjSv\%), and error in adjusted save percentage (Err), respectively.  By adjusted save percentage we mean the league average save percentage plus DiffSv\%, the difference in Sv\% and ExpSv\%.  A goalie's adjusted save percentage can be thought of as a save percentage that has been adjusted for the quality of shots he faced, and indicates what the goalie's save percentage would have been if he faced the league average quality of shots.

Last year's Vezina trophy finalists, Tim Thomas, Pekka Rinne, and Roberto Luongo, are all listed here.  Thomas, who won the award, has a sizeable lead, and was also the league leader in adjusted save percentage.  These results suggest that Henrik Lundqvist had a better season than Rinne and Luongo.  Lundqvist had a lower save percentage than Rinne and Luongo, but faced more difficult shots, according to ExpSv\%.  His DiffGA and adjusted save percentage are higher than those of Rinne and Luongo, indicating that he performed better relative to the quality of shots he faced.

We should note that a goalie's performance can fluctuate a fair amount from year to year, so we should be careful not to draw conclusions based on one season's worth of data.  In light of this, we also give the top 5 goalies in $2008$-$2011$ in Table $\ref{goalies-goals-3yrs}$.  
\begin{table}[h!]
\begin{center}
\caption{The top 5 goalies in goals prevented in $2008$-$2011$.}
\label{goalies-goals-3yrs}
{\footnotesize
\begin{tabular}{lrrrrrrrrrr}
  \addlinespace[.3em] \toprule 
Goalie & ExpGA & GA & Goals & GErr & ShotA & Sv\% & ExpSv\% &  AdjSv\% & Err \\ 
  \midrule 
Henrik Lundqvist & 581 & 461 & 120 & 0.3 & 5818 & 0.921 & 0.900 &  0.933 & 0.0001 \\ 
  Tim Thomas & 374 & 301 & 73 & 0.2 & 4411 & 0.932 & 0.915 &  0.928 & 0.0001 \\ 
  Jonas Hiller & 410 & 347 & 63 & 0.3 & 4352 & 0.920 & 0.906 &  0.927 & 0.0001 \\ 
  Tomas Vokoun & 458 & 412 & 46 & 0.3 & 5451 & 0.924 & 0.916 &  0.920 & 0.0001 \\ 
  Roberto Luongo & 432 & 388 & 44 & 0.3 & 4901 & 0.921 & 0.912 & 0.921 & 0.0001 \\ 
   \bottomrule 
\end{tabular}
}
\end{center}
\end{table}
In that table, Lundqvist is the top goalie by a sizeable margin.  However, we should also note that Alan Ryder found evidence of rink bias in the reporting of shot distance for shots taken at Madison Square Garden \cite{ryder-shot-quality-recall}.  Lundqvist's adjusted save percentage statistics are skewed by this rink bias.  
\begin{table}[h!]
\begin{center}
\caption{The top 5 goalies in goals prevented in $2008$-$2011$ (away games only).}
\label{goalies-away-games-only-3yrs}
{\footnotesize
\begin{tabular}{lrrrrrrrrrr}
  \addlinespace[.3em] \toprule 
Goalie & ExpGA & GA & Goals & GErr & ShotA & Sv\% & ExpSv\% &  AdjSv\% & Err \\ 
  \midrule 
Tim Thomas & 195 & 152 & 43 & 0.2 & 2298 & 0.934 & 0.915 &  0.929 & 0.0001 \\ 
  Henrik Lundqvist & 248 & 218 & 30 & 0.2 & 2825 & 0.923 & 0.912 &  0.920 & 0.0001 \\ 
  Craig Anderson & 225 & 196 & 29 & 0.2 & 2449 & 0.920 & 0.908 &  0.922 & 0.0001 \\ 
  Tomas Vokoun & 223 & 194 & 29 & 0.2 & 2514 & 0.923 & 0.911 &  0.921 & 0.0001 \\ 
  Cam Ward & 247 & 220 & 27 & 0.2 & 2748 & 0.920 & 0.910 &  0.920 & 0.0001 \\ 
   \bottomrule 
\end{tabular}
}
\end{center}
\end{table}
In Table $\ref{goalies-away-games-only-3yrs}$, we give results similar to those in Table $\ref{goalies-goals-3yrs}$ but for away games only, in an attempt to remove the affects of scoring bias in the goalie's home rink.  In that table, Lundqvist is second to Thomas.  We note that in an attempt to remove the effects of rink scoring bias, we will continue using performance in away games only throughout the rest of Section \ref{teamperformance}.

Finally, one might be interested to see how different adjusted save percentage can be from save percentage.  If adjusted save percentage were not much different than save percentage, then adjusted save percentage would not have much value.
\begin{table}[h!]
\begin{center}
\caption{The 5 goalies with the biggest difference in Sv\% and adjusted Sv\% (away games only).}
\label{goalies-change}
{\footnotesize
\begin{tabular}{lrrrrrrrrrrr}
  \addlinespace[.3em] \toprule 
Goalie & ExpGA & GA & Goals & GErr & ShotA & Sv\% &  AdjSv\% & Err & Change \\ 
  \midrule 
Tim Thomas & 195 & 152 & 43 & 0.2 & 2298 & 0.934 &   0.929 & 0.0001 & $-$0.005 \\ 
  Martin Brodeur & 181 & 188 & $-$7 & 0.2 & 2107 & 0.911 &   0.906 & 0.0001 & $-$0.004 \\ 
  Dwayne Roloson & 224 & 215 & 9 & 0.2 & 2401 & 0.910 &   0.913 & 0.0001 & 0.003 \\ 
  Niklas Backstrom & 196 & 177 & 19 & 0.2 & 2242 & 0.921 &   0.918 & 0.0001 & $-$0.003 \\ 
  Ryan Miller & 237 & 217 & 20 & 0.2 & 2710 & 0.920 &   0.917 & 0.0001 & $-$0.003 \\ 
   \bottomrule 
\end{tabular}
}
\end{center}
\end{table}
In Table \ref{goalies-change}, we give the 5 goalies who have the largest difference between their save percentage and adjusted save percentage during away games only.  These goalies would be the goalies whose value changes the most when using weighted shots instead of shots.  According to the ``Change'' column in this table, Dwayne Roloson would be undervalued when considering his unadjusted save percentage alone, while the others are overvalued.  We note that a change of about 0.005 over 2000 shots (roughly the number of shots that a top starting goaltender would face during one season) corresponds to about 10 goals, or roughly 2 wins if we use the conversion that 6 goals is approximately equivalent to 1 win.

\subsection{Reliability and Predictive Power of Adjusted Save Percentage}\label{section-adj-save}

In Section \ref{logistic}, we showed that, on the level of a single shot, including detailed information about that shot is beneficial for determining whether or not the shot will become a goal.  The idea behind using adjusted save percentage to explain a goalie's performance is not only intuitively appealing, but is also justified by the performance of the logistic regression model at the level of a single shot.   

Most of the criticism of shot quality models has been that shot quality was never shown to be very consistent or indicative of a goalie's or a team's ``true talent''.  For example, it was shown by Gabriel Desjardins that teams do not have much, if any, control over the distance of the shots they allow \cite{gabe-shot-distance}, an indication that a team can not control the quality of shots it allows very well.  

In light of these observations, we would like to determine if any of our advanced statistics are any more consistent or predictive than  traditional statistics.  For example, we would like to look for some statistical evidence that adjusted save percentage is more consistent than, or a better indicator of future performance than, (unadjusted) save percentage.  If not, then we can only use adjusted save percentage as a better way to evaluate or explain past performance, not as a better way to quantify a goalie's ability or predict future performance.  Ideally, adjusted save percentage could be used to both explain and predict, and this section is devoted to answering whether it can be used for the latter.  We do a similar analysis of some advanced team statistics in Section \ref{section-team-stats}.

\begin{figure}[h!]
    \centering

 \includegraphics[width=.30\linewidth]{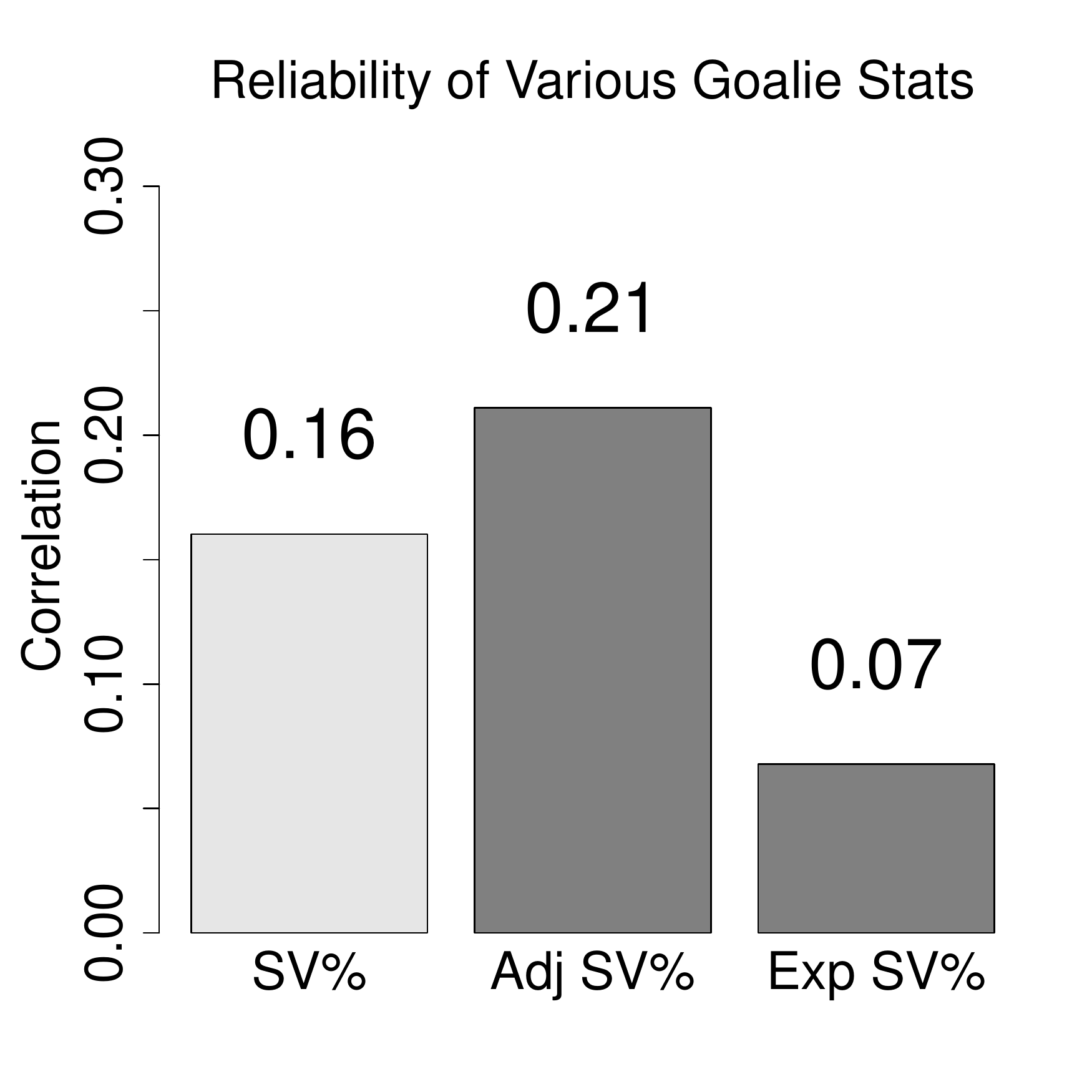} 
 \includegraphics[width=.30\linewidth]{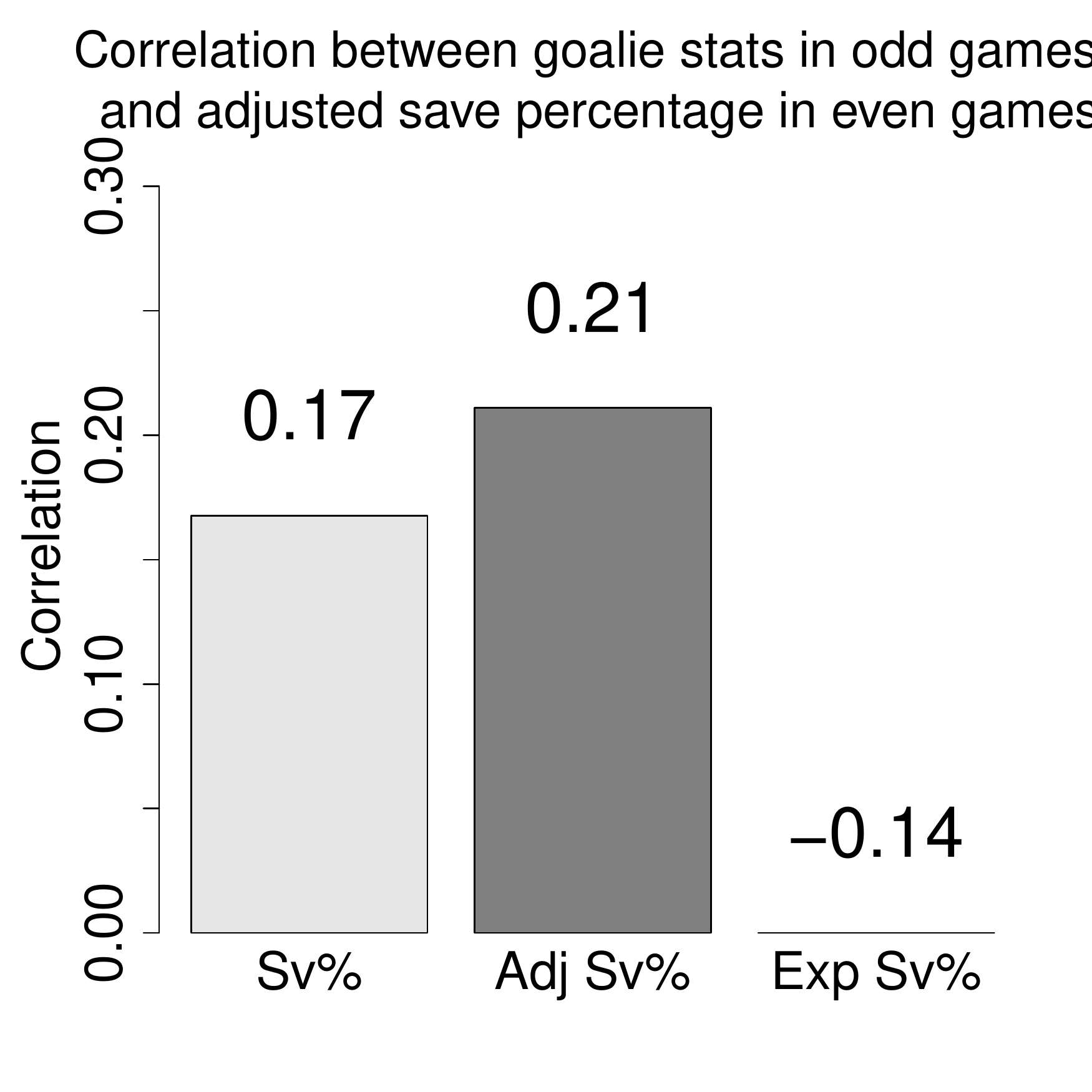} 
 \includegraphics[width=.30\linewidth]{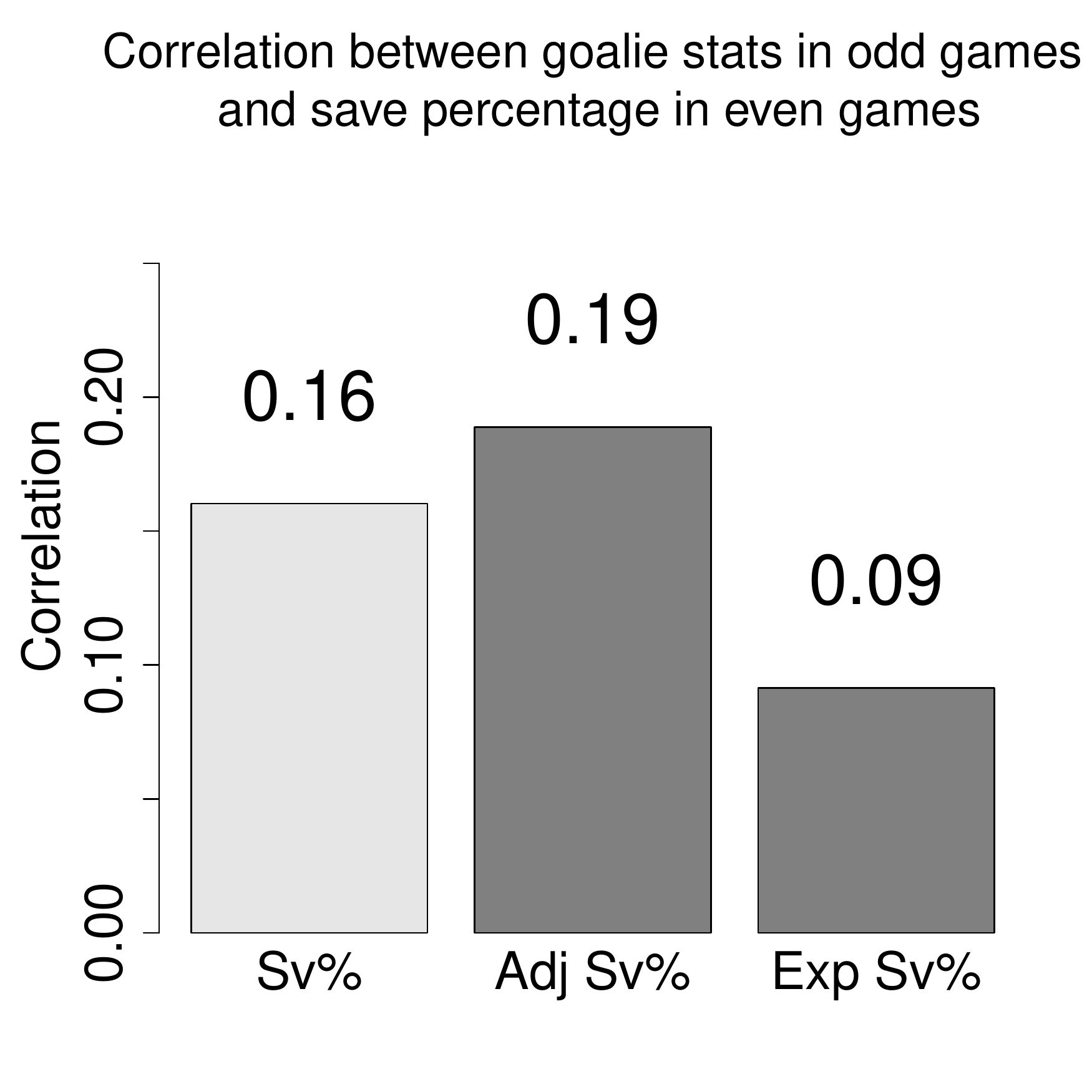} 
    \caption{(Left) Split-half Reliability of save percentage, adjusted save percentage, and expected save percentage. (Middle) Correlation between statistics in odd games and adjusted save percentage in even games.  (Right) Correlation between statistics in odd games and save percentage in even games.  These statistics are for even strength situations and are for goalies who faced a minimum of 1,000 shots during those situations in away games only.  Here and throughout this paper, dark grey will be used to indicate advanced statistics based on weighted shots.}
        \label{goalie-reliability}
    \end{figure}

In the left of Figure $\ref{goalie-reliability}$, we give the split-half reliability of save percentage, adjusted save percentage, and expected save percentage, for goalies with more than $1,000$ shots against at 5-on-5, in away games only, during the 2008-09, 2009-10, and 2010-11 seasons.  
    This figure suggests that adjusted save percentage is slightly more consistent than save percentage.  Notice also that a goalie's expected save percentage, an indication of the quality of shots he faces, is not very consistent from one half season to another.  This seems to indicate that the defense in front of a goalie is not very consistent in terms of the quality of shots it allows.

Reliability is nice to have, but our goal is to find the statistics in one half of a season that are the best indicators of a goalie's performance in the other half of the season.  Since adjusted save percentage is better than save percentage as a measure of a goalie's past performance, then one goal could be to find the statistics in one half of a season that are most correlated with adjusted save percentage in the second half of the season.  

In the middle of Figure $\ref{goalie-reliability}$, we give the correlation between some goalie statistics in odd games and adjusted save percentage in even games.  These results suggest adjusted save percentage may be a slightly better indicator of future performance than save percentage.   It also suggests that this version of adjusted save percentage may not be as independent of expected save percentage as we had hoped.  Upon further inspection, one data point is having a significant impact on this correlation, and if we were to remove that data point, the correlation is -0.02, much closer to what we would have expected.  

Finally, in the right of Figure \ref{goalie-reliability}, we give correlations between goalie statistics in odd games and save percentage in even games.  According to these results, adjusted save percentage could be a better predictor of save percentage than save percentage itself.  Of course, with all of these correlations, adjusted save percentage is only slightly higher than save percentage, and we should be careful not to draw firm conclusions from these results.  We have not found the differences in these correlations to be statistically significant.  Even if we were certain the correlations for adjusted save percentage were higher, the difference would likely not be very large.  

We stress that we have used performance in away games only, since rink bias has a fairly big impact on the results.   We illustrate the difference between the correlations when using all games and correlations when using away games only in Appendix \ref{app}. 
        
        \subsection{Reliability and Predictive Power of Team Statistics}\label{section-team-stats}
         We now explore the benefits of using statistics based on weighted shots (team adjusted save percentage, team expected save percentage, team adjusted shooting percentage, team weighted shots per 60 minutes for and against, etc.) to analyze team offensive and defensive performance.   

\begin{figure}[h]
                        \centering

 \includegraphics[width=.40\linewidth]{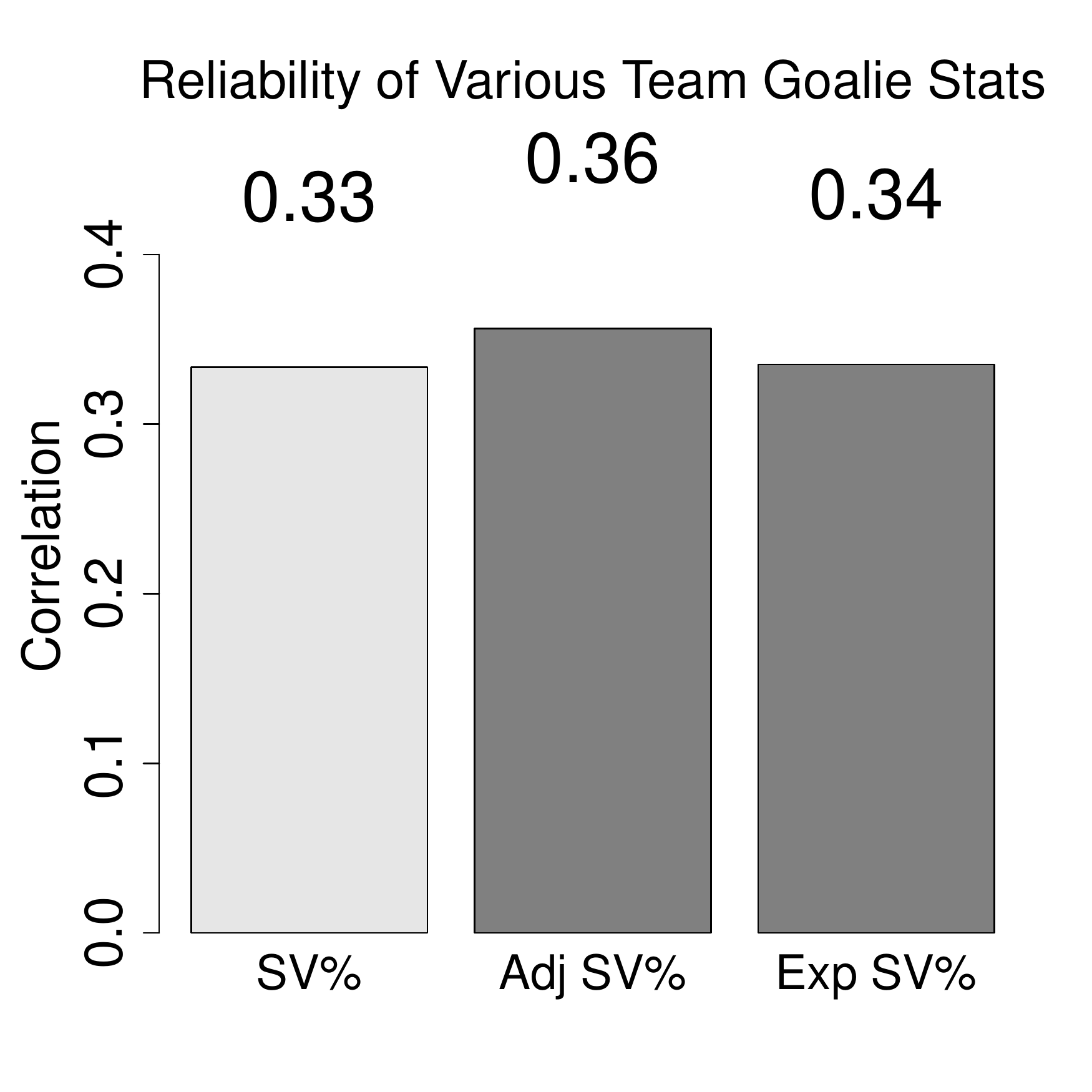}
 \includegraphics[width=.40\linewidth]{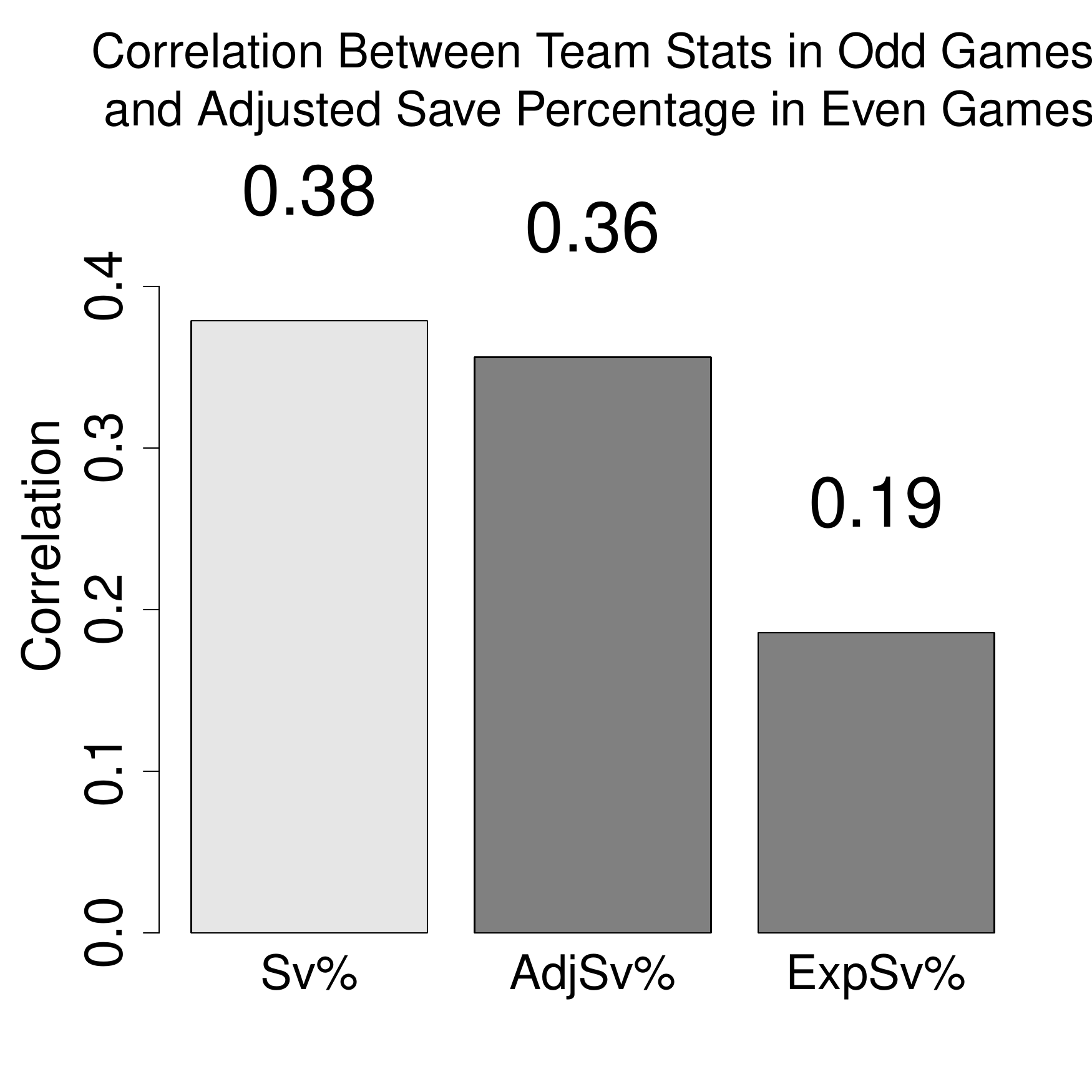} 
                        \caption{(Left) Split-half reliability of various traditional and advanced team statistics at even strength.  (Right) Correlation between team statistics in odd games and adjusted save percentage in even games. In both cases, only data from away games was used.}
                        \label{team-save}
                        \end{figure}

            In the left half of Figure $\ref{team-save}$, we show the reliability of various team goaltending statistics.

             We also give the correlation between some team statistics in odd games and adjusted save percentage in even games.  Adjusted save percentage does not seem to be a better predictor than save percentage at the team level, although the difference in correlations is once again not large and not statistically significant.  We get similar results for team shooting percentage, as shown in Appendix \ref{app}. Once again, the differences in the correlations for adjusted shooting percentage and shooting percentage are not significant. 

\begin{figure}[h!]
    \centering
\includegraphics[width=.45\linewidth]{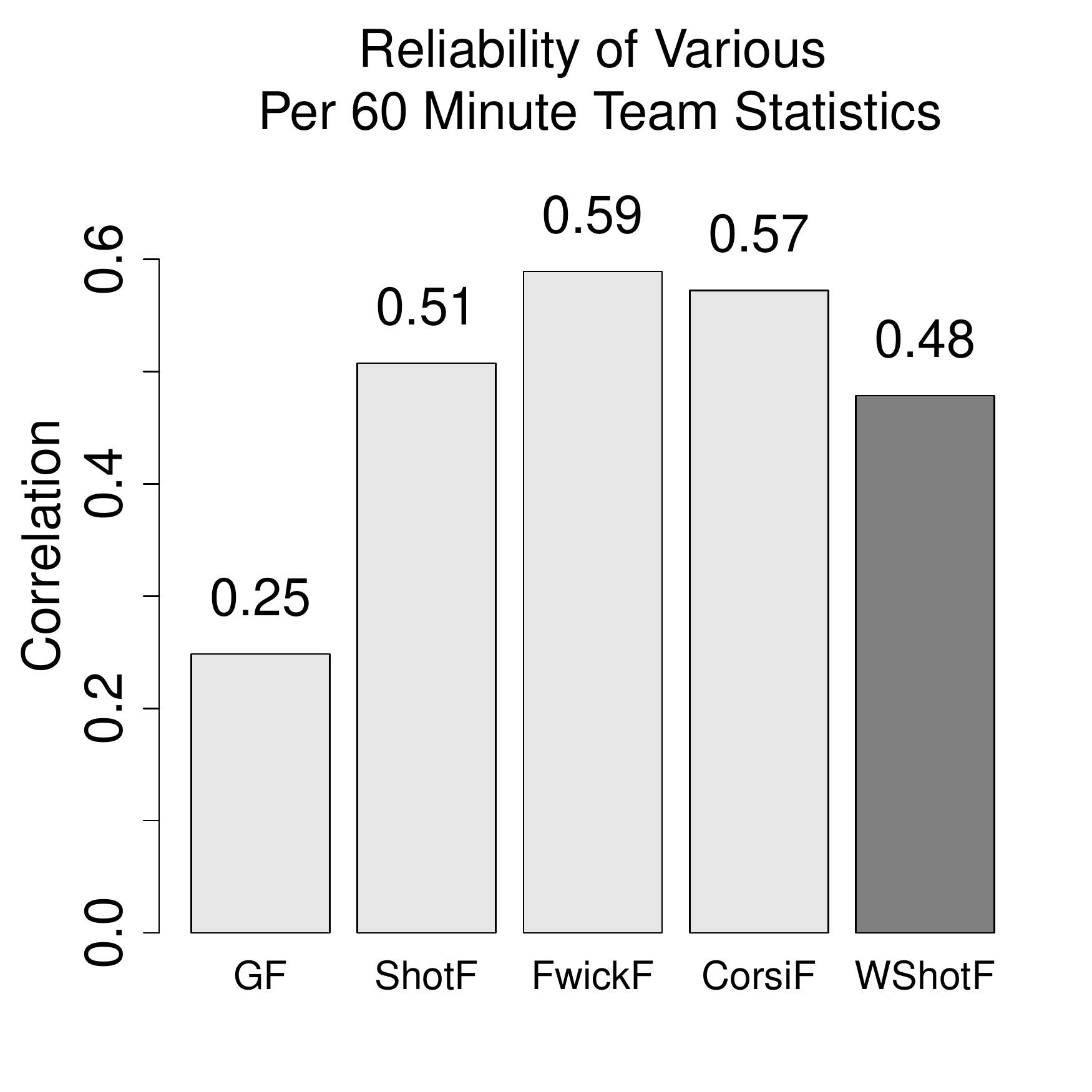}
\includegraphics[width=.45\linewidth]{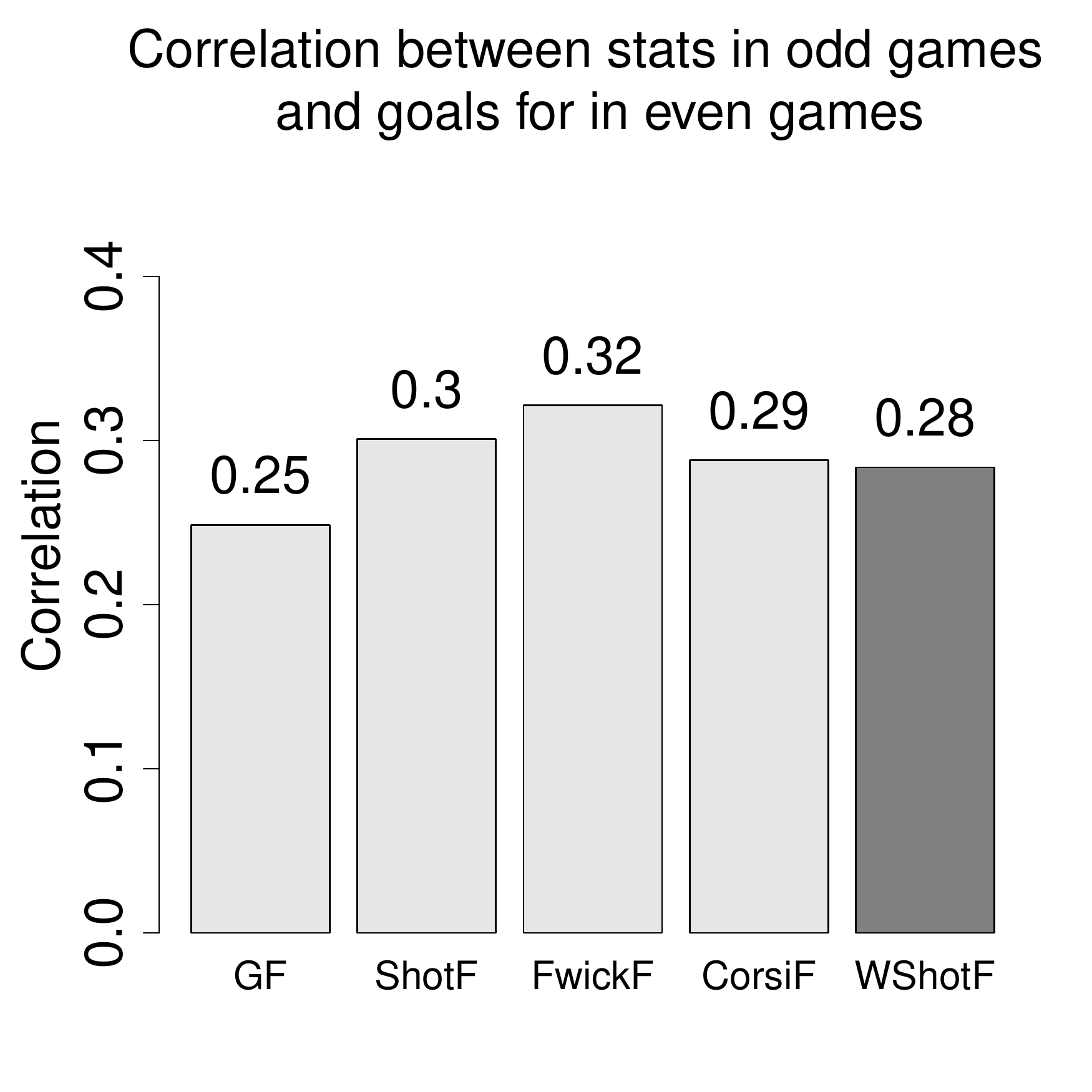}
\caption{(Left) Split-half reliability of various traditional and advanced per 60 minutes statistics at even strength.  (Right) Correlation between team statistics in odd games and goals for per 60 minutes in even games.  In both cases, only data from away games was used.}
\label{team-GF60}
    \end{figure}

We also get similar results for weighted shots per 60 minutes at the team level.  In Figure \ref{team-GF60}, we see that weighted shots per 60 minutes is much more consistent than goals per 60 minutes, but is not more consistent than shots, Fenwick (shots$+$missed shots), or Corsi (shots$+$missed shots$+$blocked shots) per 60 minutes.  We also find no evidence that weighted shots per 60 minutes outperforms shots, Fenwick, or Corsi per 60 minutes as predictor of goals for per 60 minutes.  
    In Appendix \ref{app}, we see the same general results for team defense.  That is, weighted shots against per 60 minutes performs better than goals against per 60 minutes, but not better that shots, Fenwick, or Corsi against per 60 minutes. 

We have not found any strong evidence that these advanced statistics based on weighted shots are more consistent or predictive than traditional statistics.  Nevertheless, weighted shots still performed well and often on par with those statistics.  So while we do not have evidence to lead us to use weighted shots as a replacement for any traditional statistics, it is reasonable to use weighted shots  in conjunction with those statistics, at least in the cases where weighted shots performs comparably well.

\section{Adjusted Plus-Minus using Weighted Shots}\label{apmw}
    In hockey, the plus-minus statistic is supposed to be a measure of a player's overall contribution to his team.  A player receives a $+1$ if he is on the ice when his team scores a goal at even strength or while short handed, and receives a $-1$ if he is on the ice when the opposing team scores a goal at even strength or while his team is on the power play.  

    While this statistic is supposed to be a measure of a player's offensive and defensive contributions to his team, there are several problems that limit its utility in measuring a player's \textit{individual} contribution.  For example, goal scoring can be fairly random, especially over less than a season's worth of games.  One option is to form a plus-minus statistic using weighted shots instead of goals.  There are typically about 10 times as many shots as goals, and as we have seen, weighted shots tend to be more consistent than goals.  One could also use shots, Fenwick, and Corsi instead of goals.

    But other problems still exist.  For example, a player's plus-minus is highly dependent on the strength of the players that he plays with.  A player who plays with two superstars will typically have a high plus-minus, regardless of his ability.  Also, the strength of a player's competition, and the zone on the ice in which a player typically starts his shifts, both affect a player's plus-minus.  For example, a defensive-minded defenseman is often used by his coach in a way that will tend to lower his plus-minus.  Such a defensemen will often play against the other team's best offensive players, and start his shifts in his own defensive zone, both of which will tend to lower his plus-minus. 
    
    Ideally, we would have a way of estimating a player's individual contribution to his team, independent of the strength of his teammates, the strength of his opponents, and the zones in which his shifts begin.  Attempts to adjust for one or more of these factors give more advanced statistics that are often called ``adjusted plus-minus''.  Several versions of adjusted plus-minus exist, and we describe two of them here.  
    
    First, we consider an example.  When Sidney Crosby, superstar forward of the Pittsburgh Penguins, is on the ice, he and his teammates scored 3.54 goals per 60 minutes.  When Crosby was off the ice, his teammates scored 2.51 goals per 60 minutes, a difference of 1.03 goals per 60 minutes.  This value could be considered the value that Crosby adds to his team, independent of the strength of his teammates.  A similar adjustment could be made for the strength of his opponents, and also the zones in which he begins his shifts.  
    
    This kind of adjusted plus-minus statistic, and other advanced statistics of the same flavor, have been developed by several people.  See, for example, \cite{fyffe-vollman}, \cite{seppa}, \cite{gabe}, \cite{deltasot}, \cite{tango-wowy}, \cite{wilson-wowy}, \cite{davidjohnson}, and \cite{coop}.  Weighted shots could be used in lieu of goals in any of these statistics.  In fact, in \cite{deltasot}, the author does exactly that.  Shots, Fenwick, and Corsi could also be used, as is often done by Desjardins at his website \cite{gabe}.
    
    We could also compute regression-based adjusted plus-minus statistics for each player.  Regression-based plus-minus statistics were first developed in basketball; see, for example, \cite{rosenbaum}, \cite{lewin}, \cite{ilardibarzilai}, \cite{eli}, \cite{sill}, and \cite{fearnhead-taylor-nba}.  Regression-based methods have also been used for hockey in \cite{apm}, \cite{apm2}, \cite{ridge}, and \cite{spm}.  
    
    In these regression-based methods, the observations are periods of time on the ice when no substitutions are made, the predictors are indicator variables for each player indicating whether or not they were on the ice during the observation, and the outcome variable is goals per 60 minutes during the observation.  See, for example, \cite{apm} for a detailed explanation.  Each coefficient is an estimate of a player's contribution to his team in terms of goals per 60 minutes, independent of all of the other players in the league, or in other words, independent of both his teammates and opponents.  If indicator variables are used to denote the zone in which each observation began, the coefficients give estimates that are independent of zone starts as well. 
    
    Instead of using goals as the outcome variable, we can use expected goals based on weighted shots.  
    One downside is that when using weighted shots, we are ignoring the shooting or finishing ability of players.  On the other hand, one upside is that we are also removing the effects of goalies, since for the most part, a goalie does not affect the quality of shots his team allows.  We could also use shots, Fenwick, and Corsi instead of goals, which was done in \cite{ridge}, for the same reasons.

    Early adjusted plus-minus models used ordinary least squares for estimating the coefficients.  For our adjusted plus-minus model in this paper, we use ridge regression similar to what was used in \cite{sill}, \cite{ridge}, and \cite{spm}.  Ridge regression is useful because of the collinearity in the data that occurs when certain teammates spend most of their time on the ice playing together.  See \cite{sill} or \cite{ridge} for more on the benefits of ridge regression.  
    
    In Table $\ref{offense}$, we list the top ten offensive players in adjusted plus-minus based on goals ($G$), along with adjusted plus-minus results based on weighted shots ($W$) from our model in this paper.  We also include adjusted plus-minus based on shots ($S$), Fenwick rating ($F$), and Corsi rating ($C$), as was done in \cite{ridge}.  Here, we are using data from the $2007$-$08$, $2008$-$09$, $2009$-$10$, and $2010$-$11$ seasons.\footnote{We note that for the $2008$-$09$, $2009$-$10$, and $2010$-$11$ seasons, we have used the results of our weighted shots model.  We do not have data for the $(x,y)$ coordinates for each shot in the $2007$-$08$ season, so for that season we have used the results of Krzywicki's model from \cite{ken2}.}
    
\begin{table}[h!]
\begin{center}
\caption{The top 10 offensive players in the NHL according to $G$.}
\label{offense}
{\footnotesize
\begin{tabular}{rlllrrrrrrrrr}
  \addlinespace[.3em] \toprule 
 & Player & Pos & Team & $G$ & $W$ & $S$ & $F$ & $C$ & $G_{\text{EV,60}}$ & $W_{\text{EV,60}}$ & $G_{\text{PP,60}}$ & $W_{\text{PP,60}}$ \\ 
  \midrule 
1 & Sidney Crosby & C & PIT & 23 & 17 & 12 & 13 & 14 & 0.83 & 0.66 & 0.98 & 0.61 \\ 
  2 & Jonathan Toews & C & CHI & 18 & 17 & 8 & 8 & 9 & 0.45 & 0.56 & 1.67 & 1.22 \\ 
  3 & Alex Ovechkin & LW & WSH & 17 & 17 & 17 & 20 & 24 & 0.46 & 0.43 & 0.87 & 0.93 \\ 
  4 & Daniel Sedin & LW & VAN & 16 & 14 & 13 & 13 & 15 & 0.47 & 0.44 & 1.11 & 0.88 \\ 
  5 & Joe Thornton & C & S.J & 16 & 14 & 11 & 11 & 15 & 0.34 & 0.25 & 1.24 & 1.40 \\ 
  6 & Nicklas Backstrom & C & WSH & 16 & 7 & 11 & 12 & 14 & 0.23 & 0.14 & 1.87 & 0.75 \\ 
  7 & Evgeni Malkin & C & PIT & 15 & 17 & 11 & 11 & 12 & 0.40 & 0.58 & 0.99 & 0.70 \\ 
  8 & Ryan Getzlaf & C & ANA & 15 & 7 & 6 & 8 & 9 & 0.31 & 0.09 & 1.75 & 0.95 \\ 
  9 & Pavel Datsyuk & C & DET & 15 & 6 & 10 & 11 & 12 & 0.53 & 0.08 & 0.77 & 0.84 \\ 
  10 & Jason Spezza & C & OTT & 13 & 5 & 7 & 8 & 9 & 0.37 & 0.13 & 1.38 & 0.64 \\ 
  10 & Henrik Sedin & C & VAN & 13 & 13 & 8 & 9 & 11 & 0.29 & 0.37 & 1.13 & 0.77 \\ 
  10 & Zach Parise & LW & N.J & 13 & 17 & 12 & 12 & 11 & 0.49 & 0.74 & 1.12 & 1.13 \\ 
   \bottomrule 
\end{tabular}
}
\end{center}
\end{table}

Sidney Crosby is the league leader in $G$ and is tied for second in $W$, despite missing several games during the past four full NHL seasons.  The other players on this list are considered among the league's elite offensive players.  Three players are tied in $10$th, but we would probably consider Zach Parise as the best of the three given his superior results in $W$, $S$, $F$, and $C$.  Interestingly, the league leader in $W$, Henrik Zetterberg, did not make this list.  Zetterberg is a top $3$ player in all of the shot based results ($W$, $S$, $F$, $C$), so one might choose to consider him as one of the top $10$ offensive players in the league.  

\section{Conclusions and Future Work}\label{conclusions}
    We formed a logistic regression model to estimate the probability that a shot will be a goal.  This model was similar to Krzywicki's models in \cite{ken1}, \cite{ken2}, \cite{ken3}, but with some additional variables that quantify the fatigue of the shooter and the time that the offense and defense had been on the ice at the time of the shot.  We used this model to create advanced statistics based on weighted shots, and analyzed the reliability and predictive power of these statistics in comparison to more traditional statistics.  We did not find any strong evidence that weighted shots outperforms shots, though weighted shots still performed well and we conclude that it is reasonable to use weighted shots in addition to, but not as a replacement for, shots.  Finally, we used weighted shots per 60 minutes in an adjusted plus-minus model for estimating the contribution of a player to his team, independent of the strength of his teammates, strength of his opponents, and the zones in which he typically begins his shifts.
    
    One model extension we will explore is the use of random effects to address possible correlations between observations in the data.  Our first effort will be to fit a random effects model with the goalie as the cluster variable.  In addition to addressing the possible concern with correlation in the current model, the random effects model offers an additional potential to determine how much difference the goalie makes.  Further, by producing posterior estimates of the random effect for each goal, we may have an alternate means by which we can identify goalies that are particularly effective or ineffective.  We can also account for rink bias in a similar way, and an improvement in this area could produce a better model at the level of a single shot.  Ideally, the resulting weighted shots statistics would be more reliable and predictive at the goalie or team level as well.
    
    Another possible modification we will make is to build separate models for even strength, power play, and shorthanded situations.  Different kinds of shots may be better in different situations, and with the current model it would be difficult to account for those differences without using several interaction terms in the model.  
    Another way to deal with interactions is to use a non-parametric regression such as the one developed in \cite{digr}, and this approach may yield improved performance.

\vskip 1cm
\appendix

\section{Results of the Logistic Regression Model}
        \begin{table}[h!]
        \begin{center}
        \caption{A summary of the results of our logistic regression model}
        \label{summary}
        {\footnotesize
        \begin{tabular}{rrrrrrl}
          \addlinespace[.3em] \toprule 
         & Coeff & Error & Odds & Z-val & P-value & Signif. \\ 
          \midrule 
        (Intercept) & $-$1.333 & 0.083 & 0.26 & $-$16.10 & $<.0001$ & *** \\ 
          Own rebound& $-$0.531 & 0.099 & 0.59 & $-$5.35 & $<.0001$ & *** \\ 
          Rebound & 0.547 & 0.063 & 1.73 & 8.73 &          $<.0001$ & *** \\ 
          Distance & $-$0.054 & 0.001 & 0.95 & $-$78.98 &  $<.0001$ & *** \\ 
          Angle & $-$0.017 & 0.000 & 0.98 & $-$37.98 &     $<.0001$ & *** \\ 
          Back & 0.687 & 0.081 & 1.99 & 8.49 & $<.0001$ & *** \\ 
          Slap & 1.411 & 0.083 & 4.10 & 16.97 & $<.0001$ & *** \\ 
          Snap & 1.135 & 0.081 & 3.11 & 14.03 & $<.0001$ & *** \\ 
          Tip & 0.803 & 0.087 & 2.23 & 9.27 & $<.0001$ & *** \\ 
          Wrist & 0.954 & 0.079 & 2.60 & 12.13 & $<.0001$ & *** \\ 
          EV44 & $-$0.328 & 0.057 & 0.72 & $-$5.81 & $<.0001$ & *** \\ 
          PP54 & 0.362 & 0.021 & 1.44 & 17.12 & $<.0001$ & *** \\ 
          PP53 & 0.929 & 0.061 & 2.53 & 15.31 & $<.0001$ & *** \\ 
          SH45 & 0.189 & 0.048 & 1.21 & 3.96 & $.0001$ & *** \\ 
          SH35 & 1.277 & 0.501 & 3.58 & 2.55 & $.0108$ & * \\ 
          Angle Change Left & 0.013 & 0.002 & 1.01 & 8.64 &$<.0001$ & *** \\ 
          Angle Change Right & 0.014 & 0.001 & 1.01 & 10.02 & $<.0001$ & *** \\ 
          Shooter fatigue & $-$0.025 & 0.001 & 0.97 & $-$46.95 & $<.0001$ & *** \\ 
          Off Time on ice & 0.022 & 0.001 & 1.02 & 32.09 & $<.0001$ & *** \\ 
          Def Time on ice& 0.001 & 0.001 & 1.00 & 1.90 & $.0576$ & . \\ 
          Scorediff & 0.031 & 0.005 & 1.03 & 6.38 & $<.0001$ & *** \\ 
          Byhome & $-$0.024 & 0.016 & 0.98 & $-$1.55 & $.121$ &   \\ 
          Reb:Angle & 0.005 & 0.002 & 1.01 & 2.65 & $.00809$ & ** \\ 
          Own:Angle & 0.007 & 0.003 & 1.01 & 1.99 & $.0460$ & * \\ 
          Tip:Angle & 0.014 & 0.001 & 1.01 & 9.58 & $<.0001$ & *** \\ 
           \bottomrule 
        \end{tabular}
        }
        \end{center}
        \end{table}

\section{Rink Bias}\label{app}

    Here we illustrate the effect that rink bias can have on the correlations in Section 5 by giving results based on all games (left column) and results based on away games only (right column). 
    
\newpage

\section*{Team Shooting} 

\begin{figure}[h!]
 \centering
 \includegraphics[width=.40\linewidth]{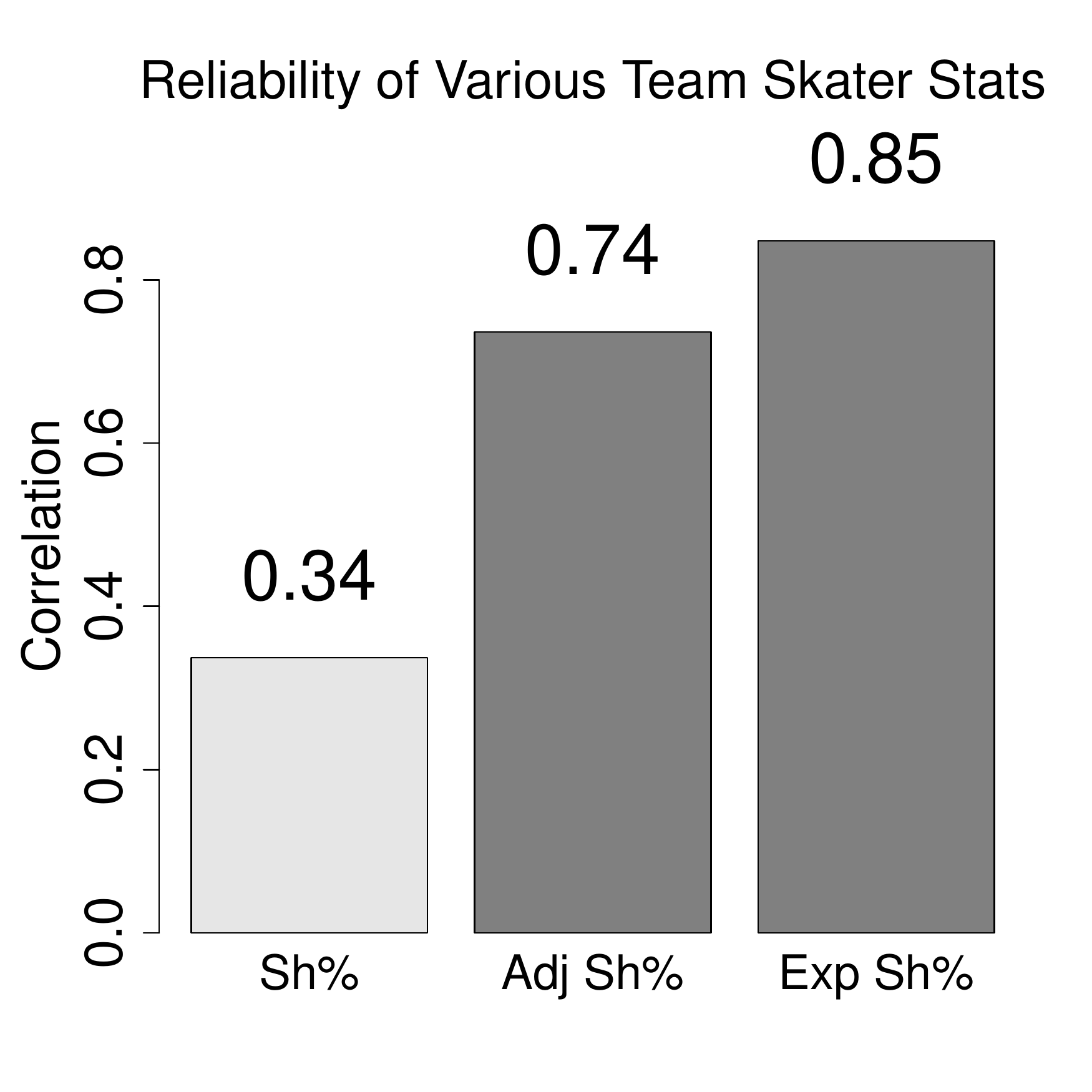}  \includegraphics[width=.40\linewidth]{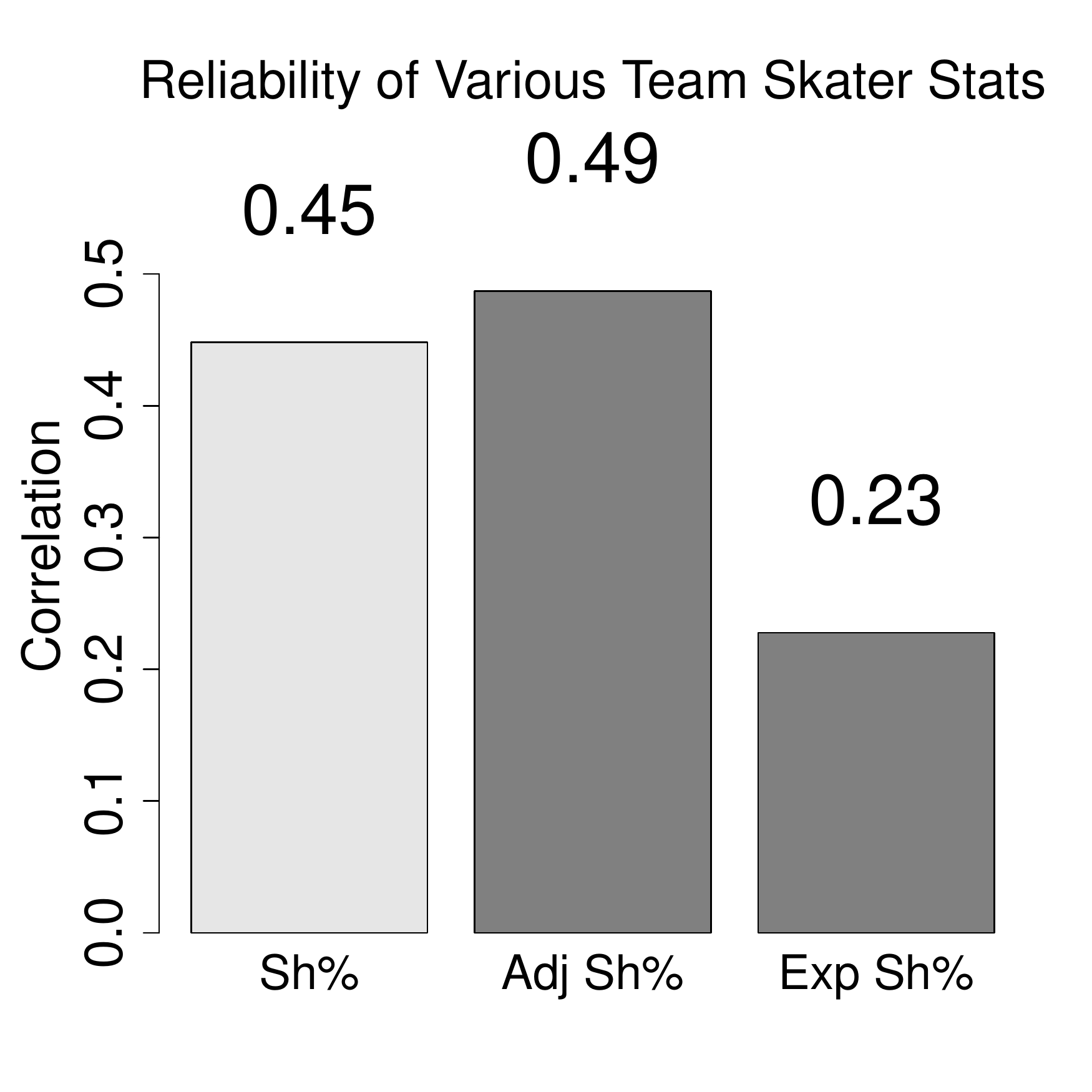} 

 \includegraphics[width=.40\linewidth]{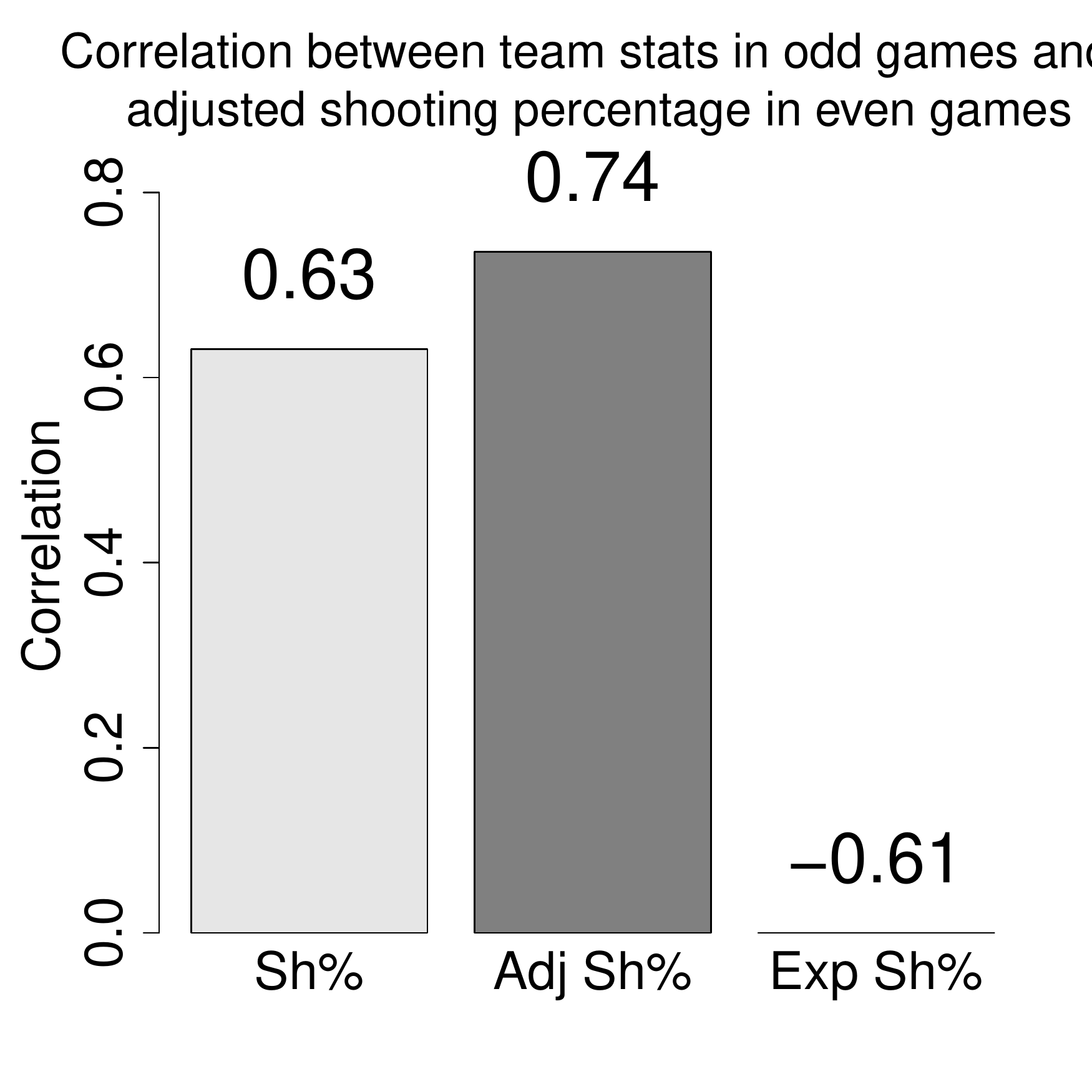}  \includegraphics[width=.40\linewidth]{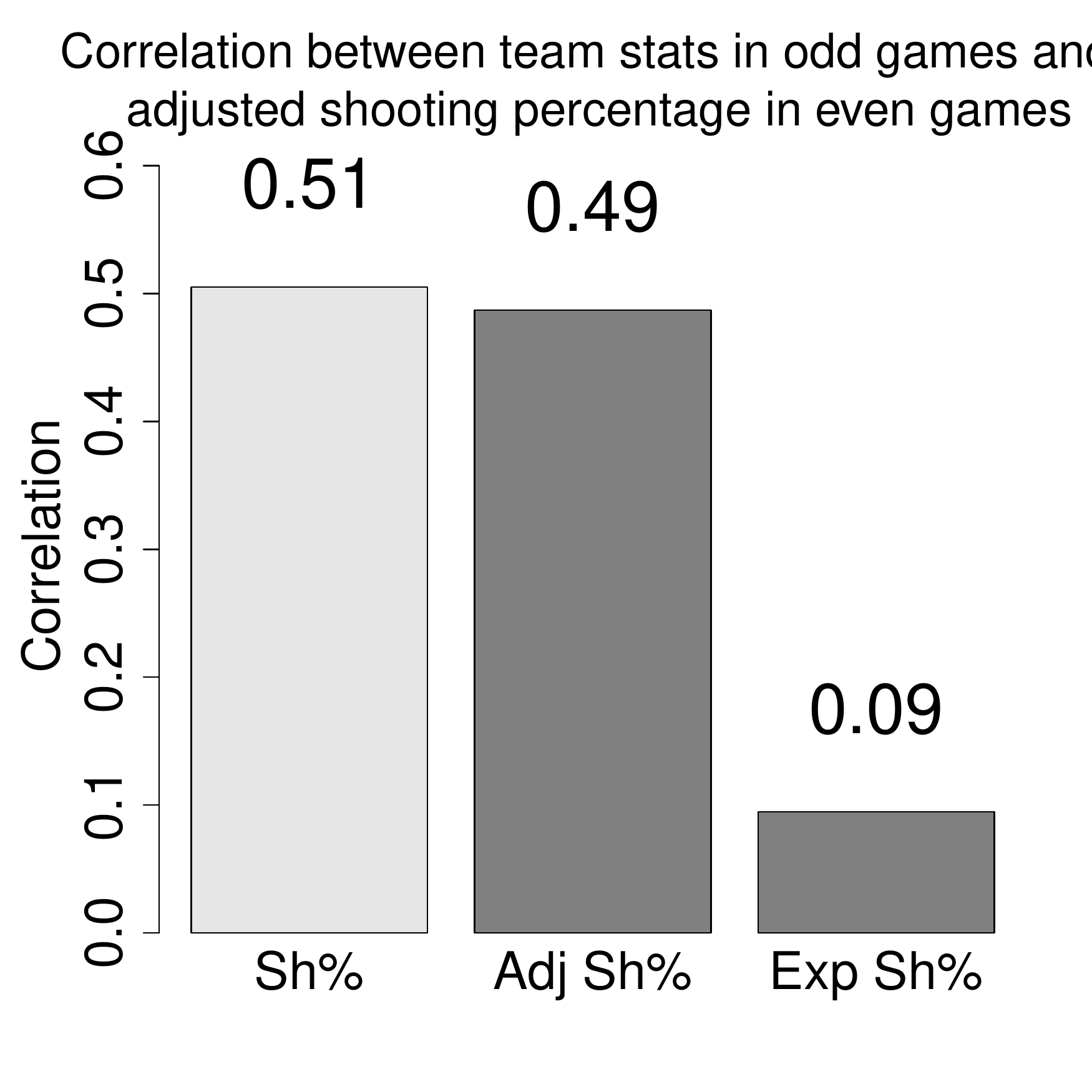} 

 \includegraphics[width=.40\linewidth]{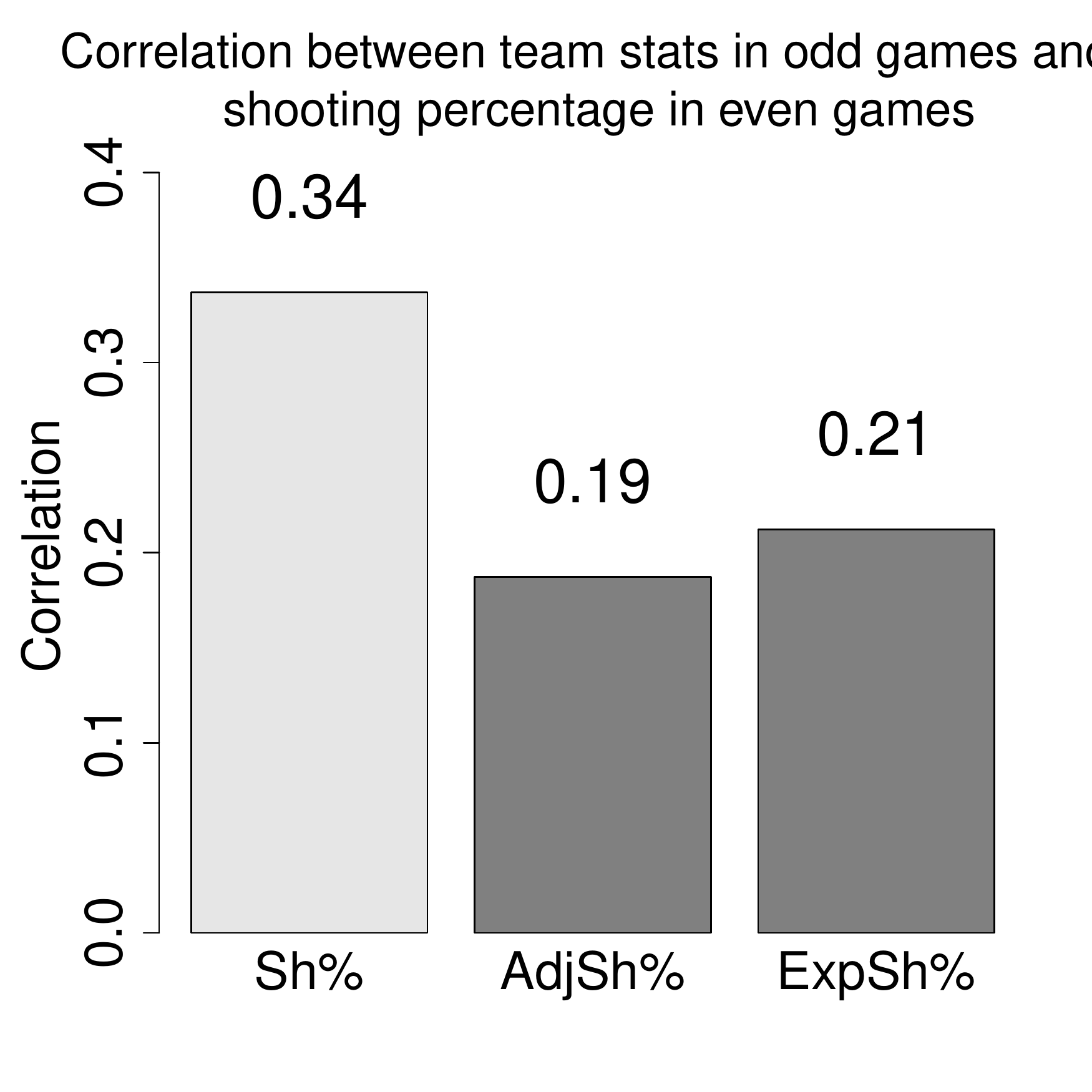}  \includegraphics[width=.40\linewidth]{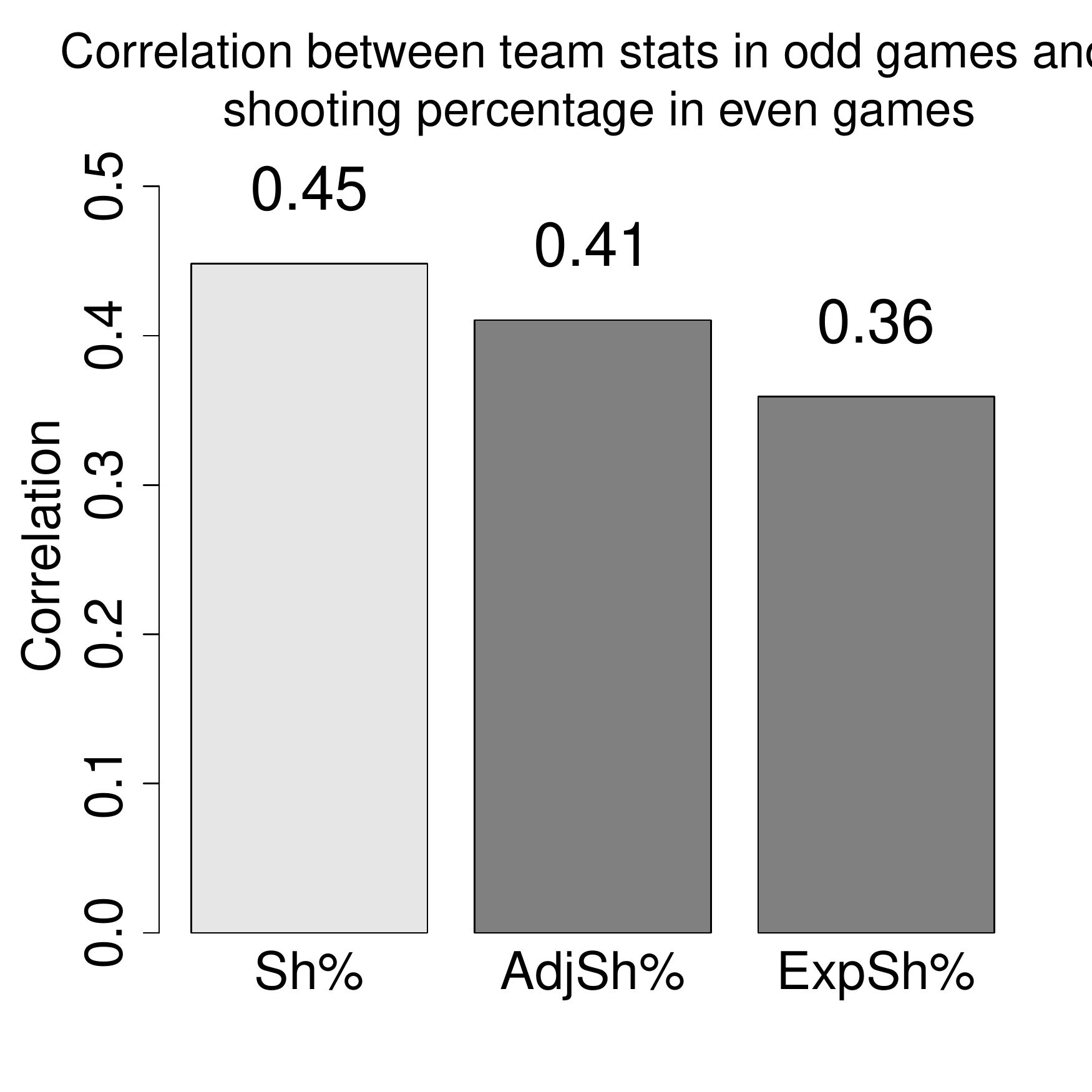} 
 \caption{Correlations for team shooting statistics}
 \label{team-shooting-with-away}
 \end{figure}

\newpage
\section*{Goalies}
\begin{figure}[h!]
 \centering
 \includegraphics[width=.40\linewidth]{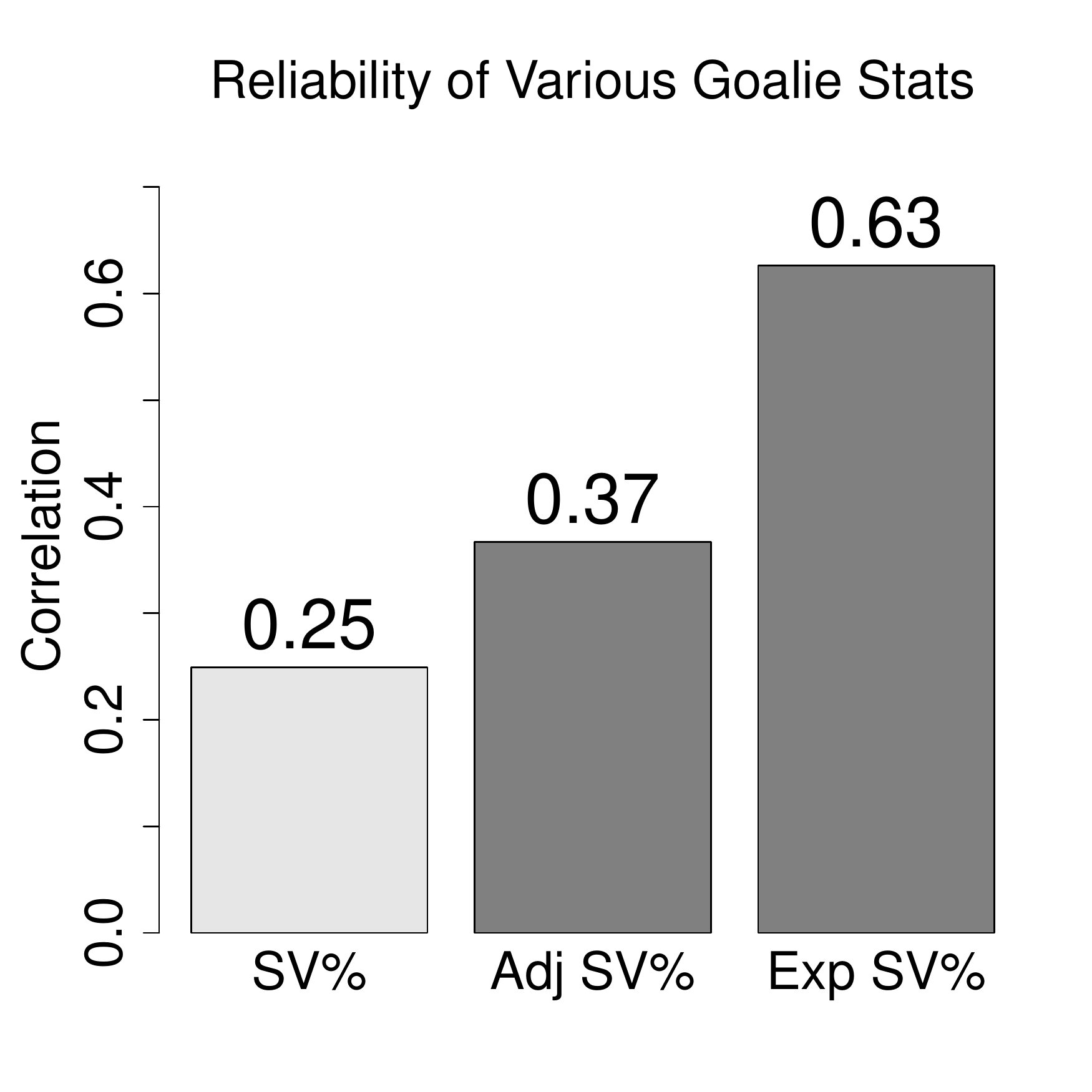} 
  \includegraphics[width=.40\linewidth]{EV55-home-goalie-rel-save.pdf} 

 \includegraphics[width=.40\linewidth]{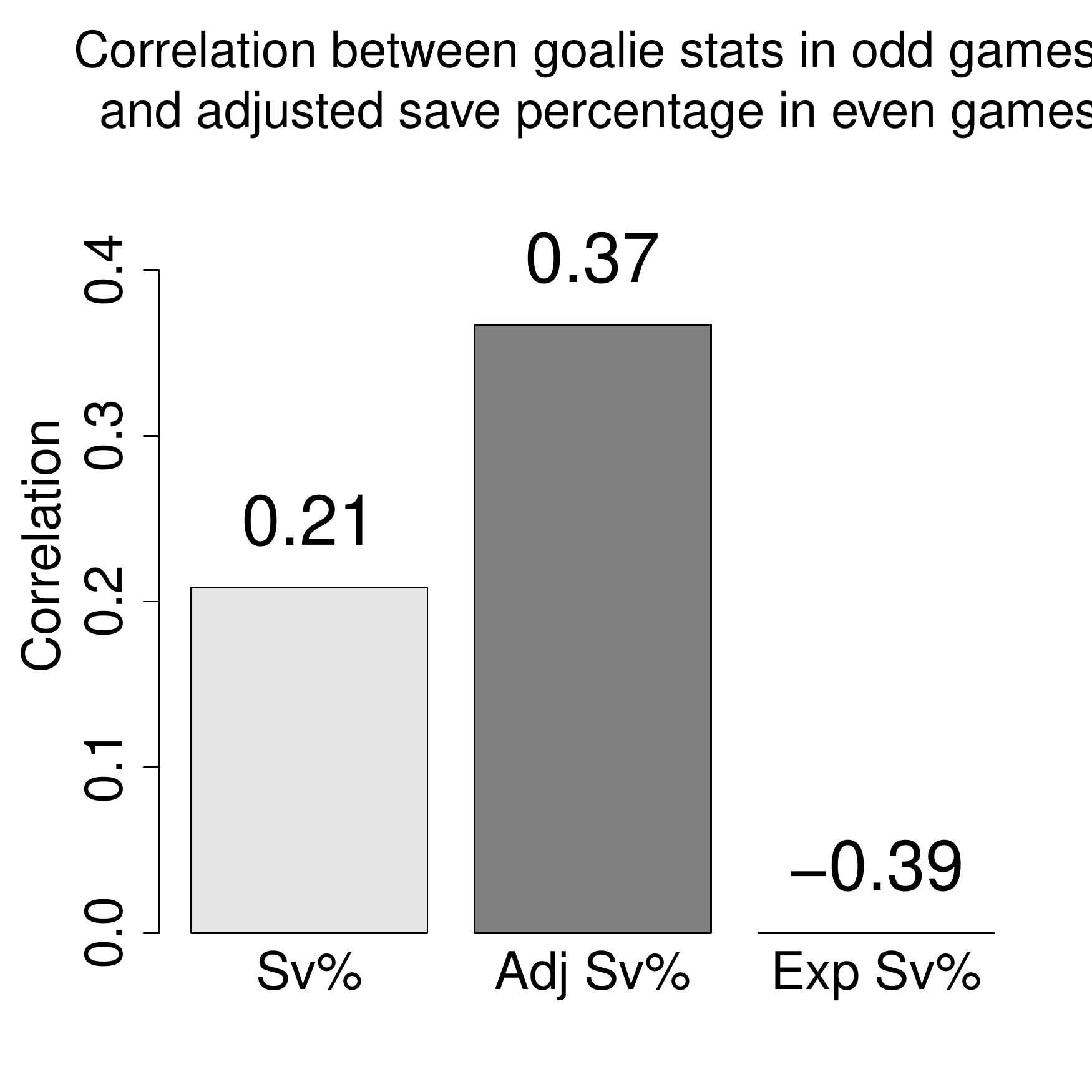} 
  \includegraphics[width=.40\linewidth]{EV55-home-goalie-pred-adjsv.pdf} 

\includegraphics[width=.40\linewidth]{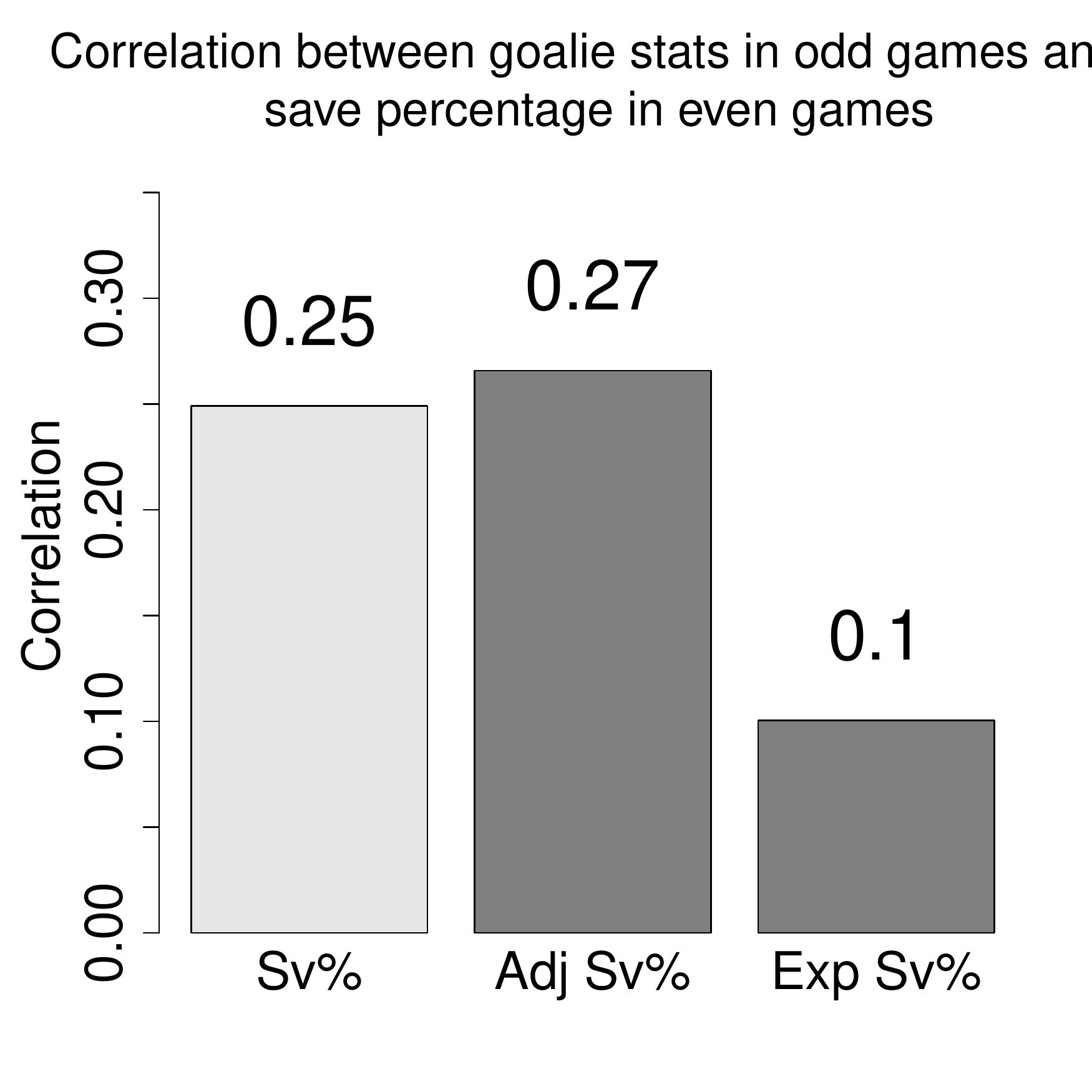} \includegraphics[width=.40\linewidth]{EV55-home-goalie-pred-save.pdf}
\caption{Correlations for goalie statistics.}
\label{goalie-save-with-away}

\end{figure}
\newpage
\section*{Team Goaltending}
\begin{figure}[h!]
 \centering
 \includegraphics[width=.40\linewidth]{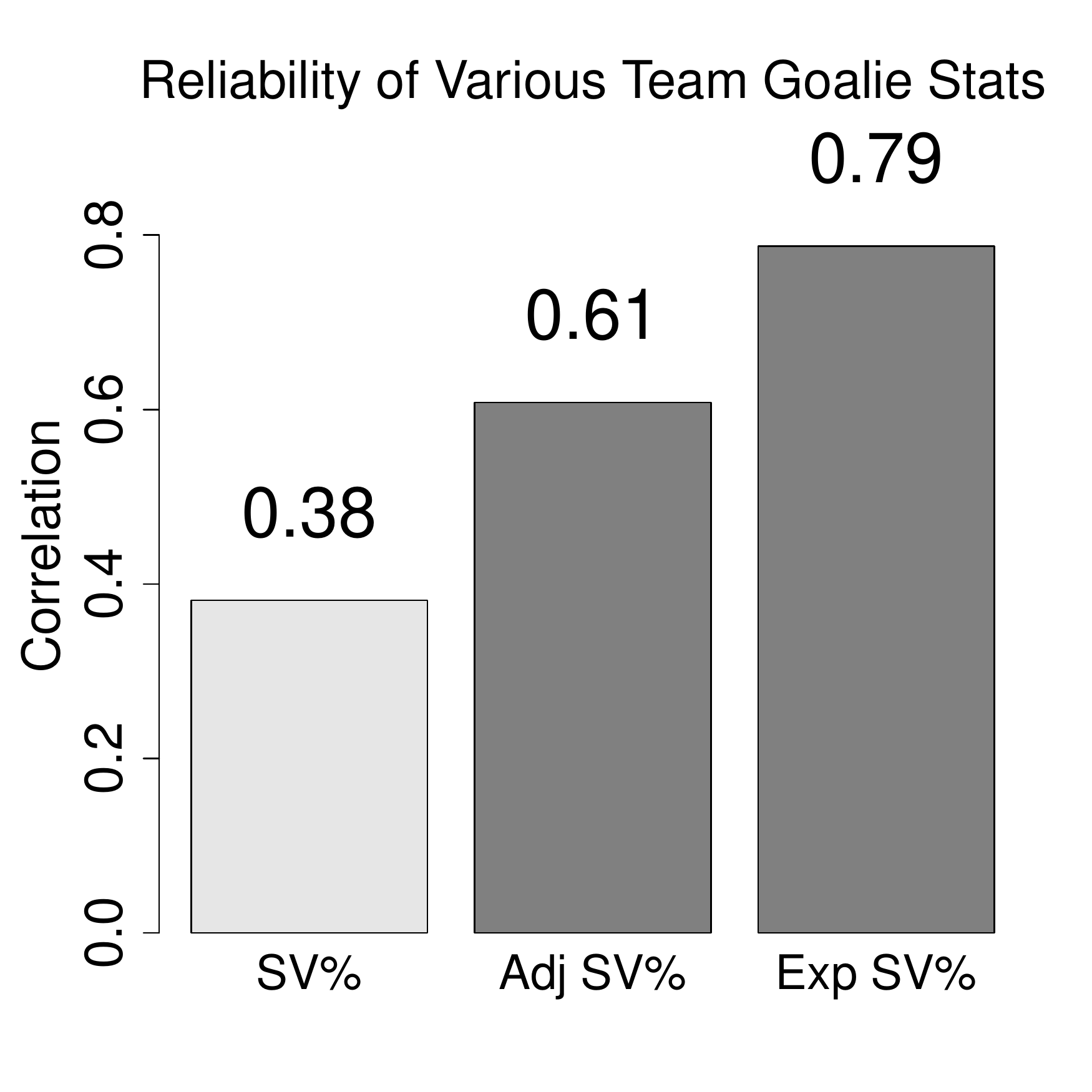}  \includegraphics[width=.40\linewidth]{EV55-home-teams-rel-save.pdf} 

 \includegraphics[width=.40\linewidth]{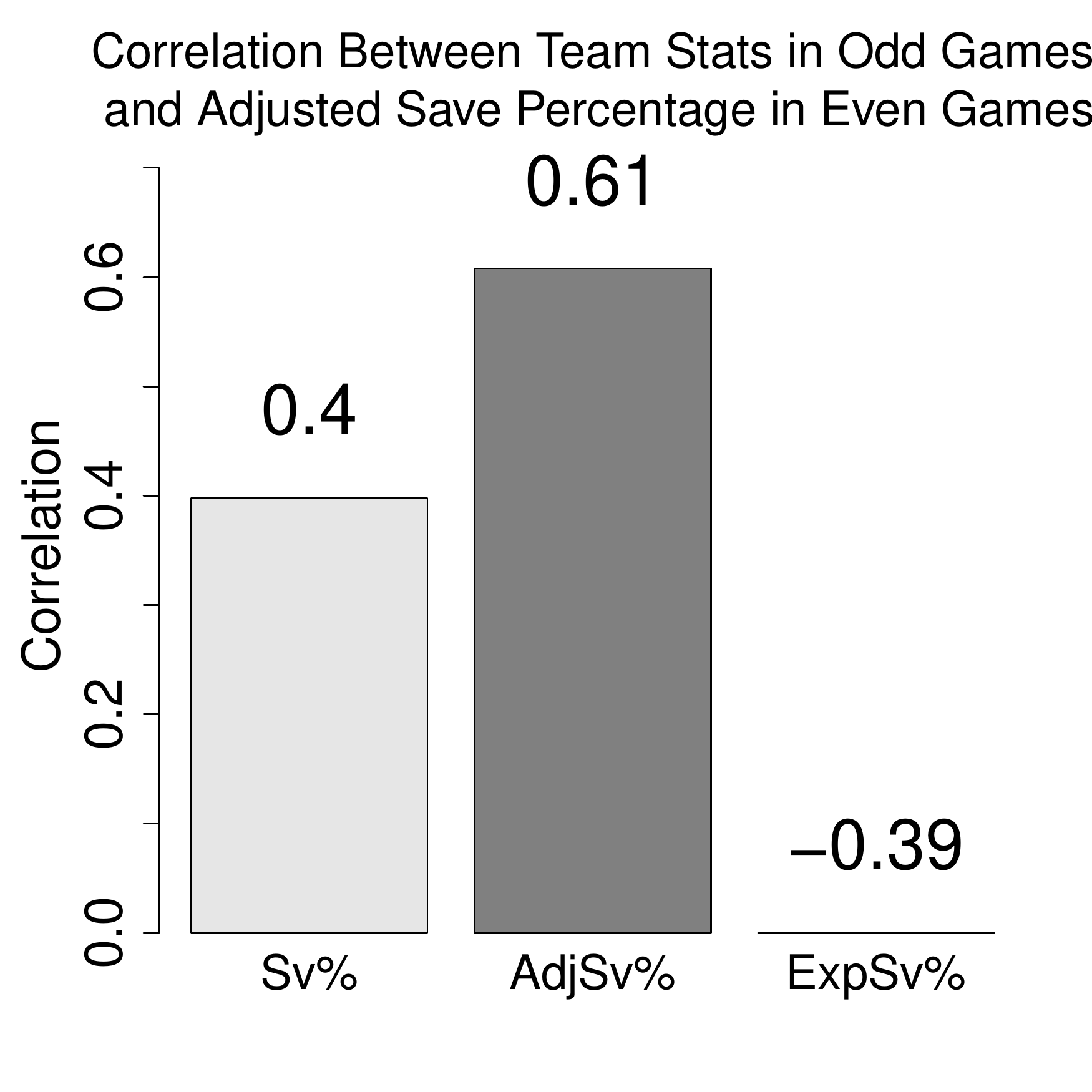}  \includegraphics[width=.40\linewidth]{EV55-home-teams-pred-adjsv.pdf}

 \includegraphics[width=.40\linewidth]{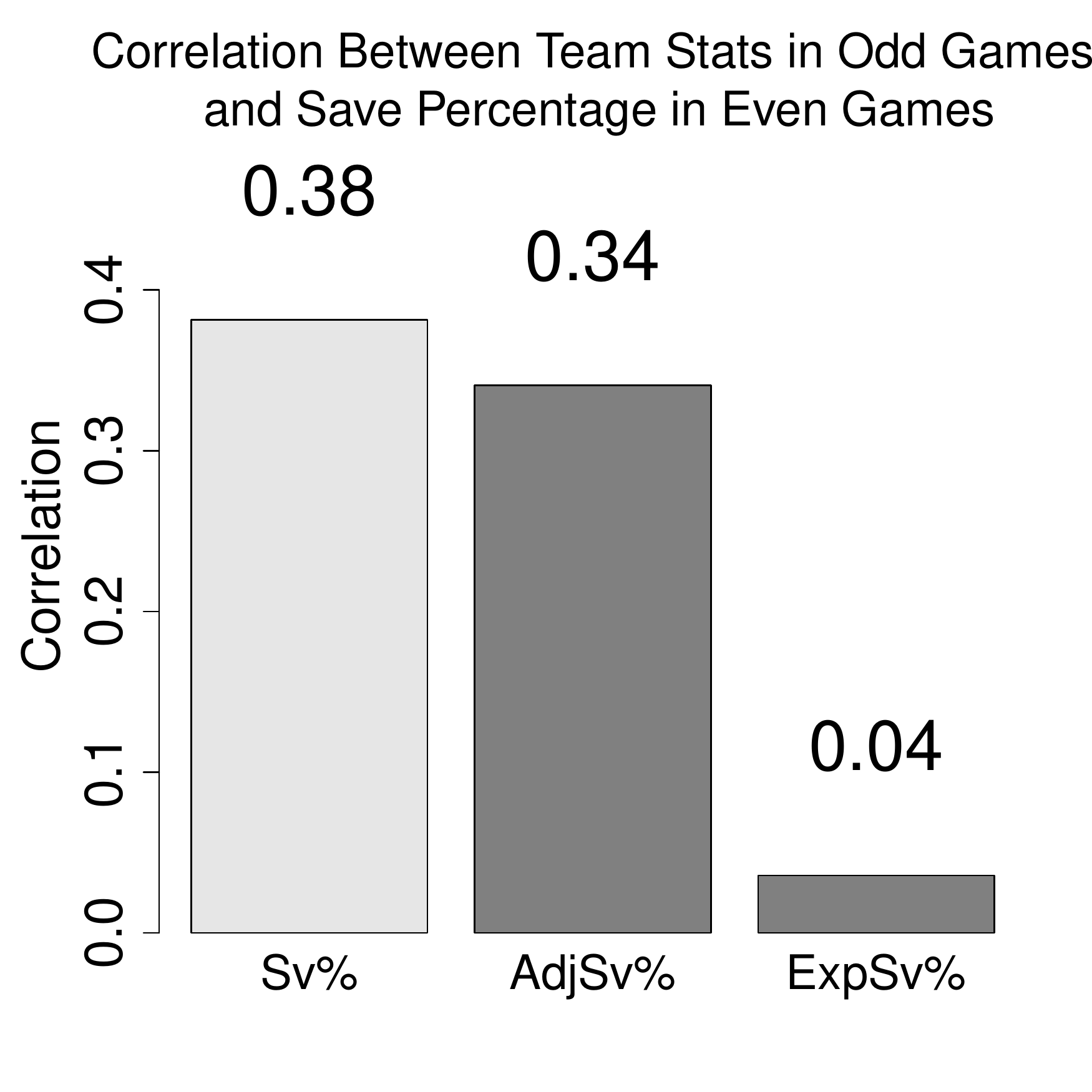}  \includegraphics[width=.40\linewidth]{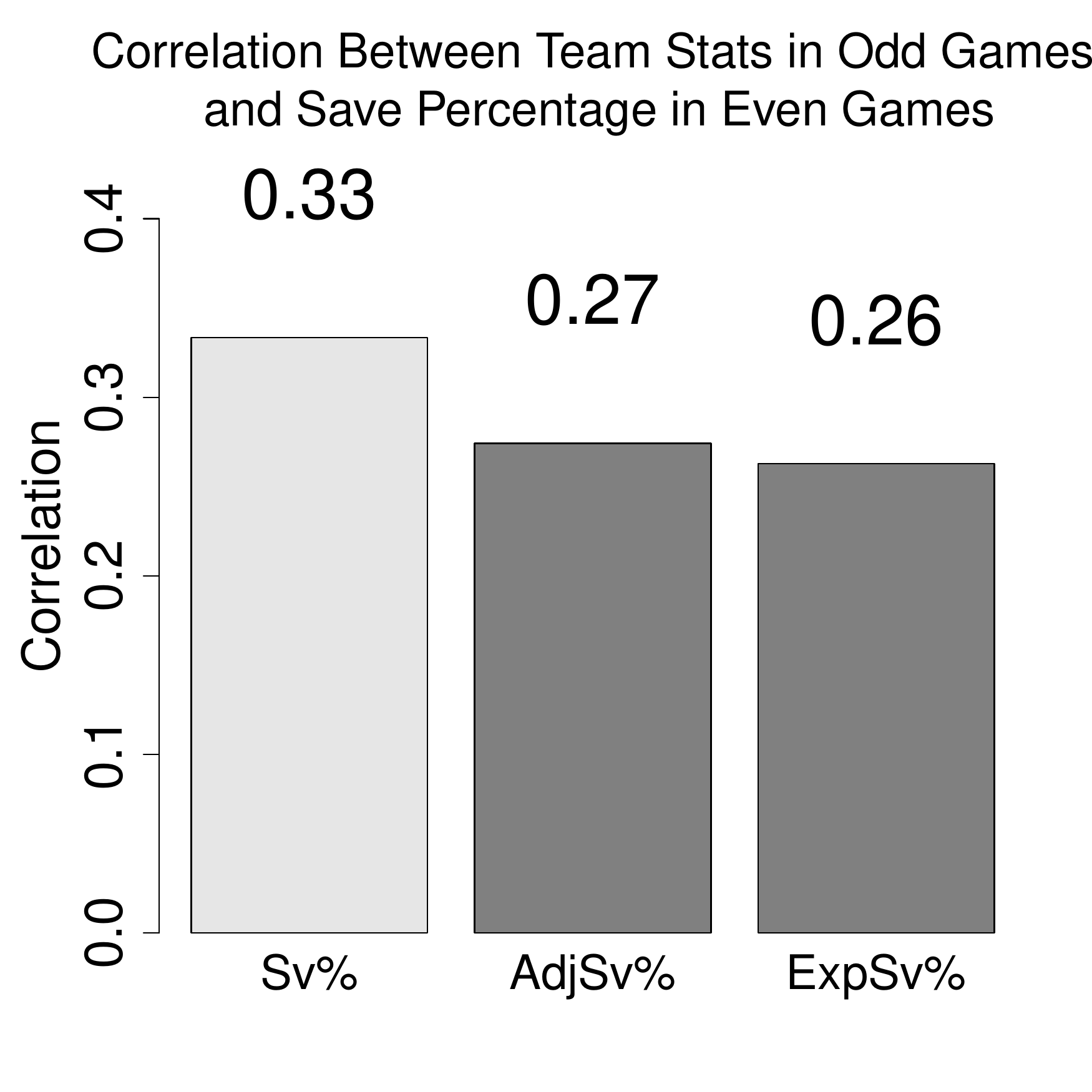} 
 \caption{Correlations for team goalie statistics.}
 \label{team-save-with-away}
 \end{figure}

\newpage 
\section*{Team Defense}\label{team-defense}
\begin{figure}[h!]
    \centering
\includegraphics[width=.40\linewidth]{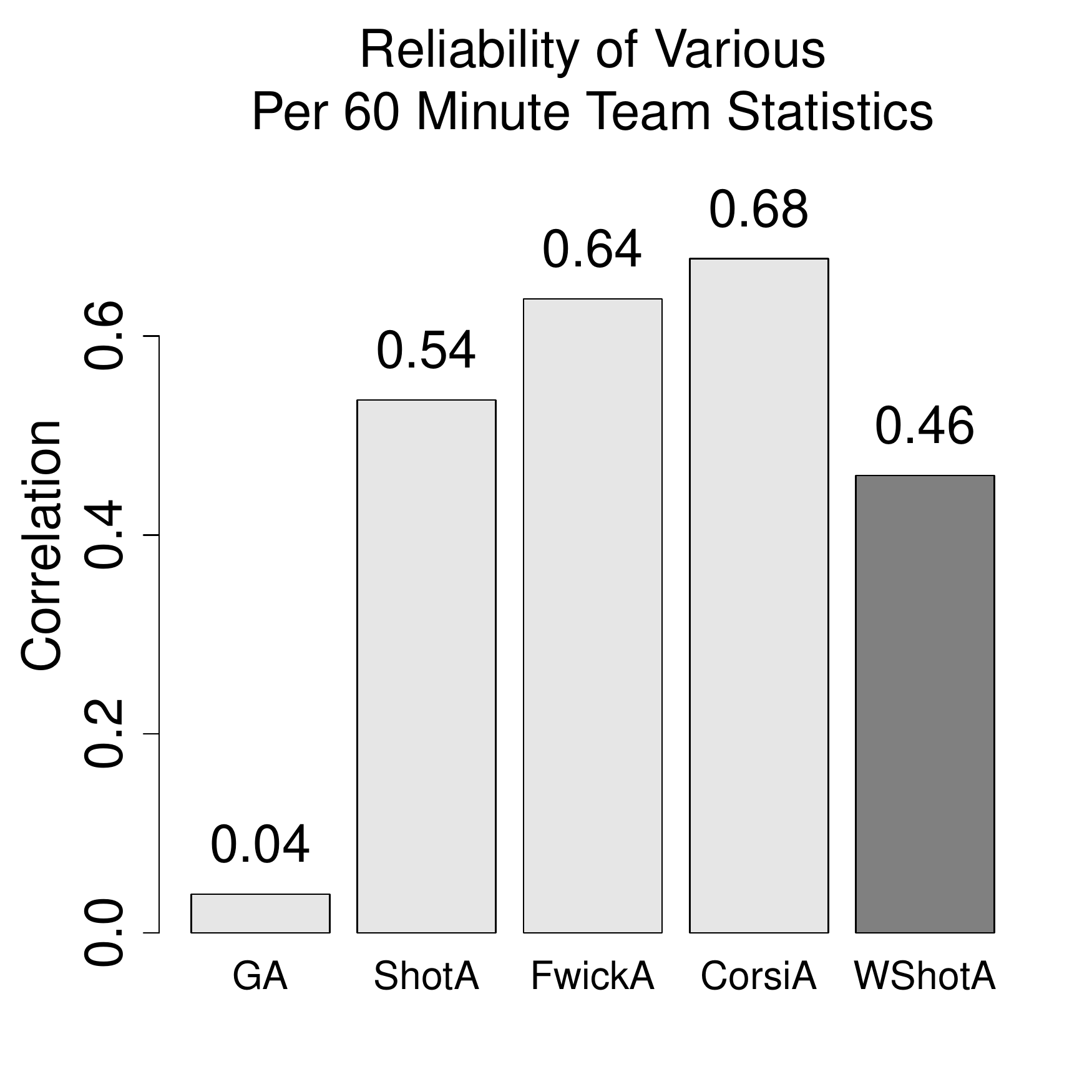}
\includegraphics[width=.40\linewidth]{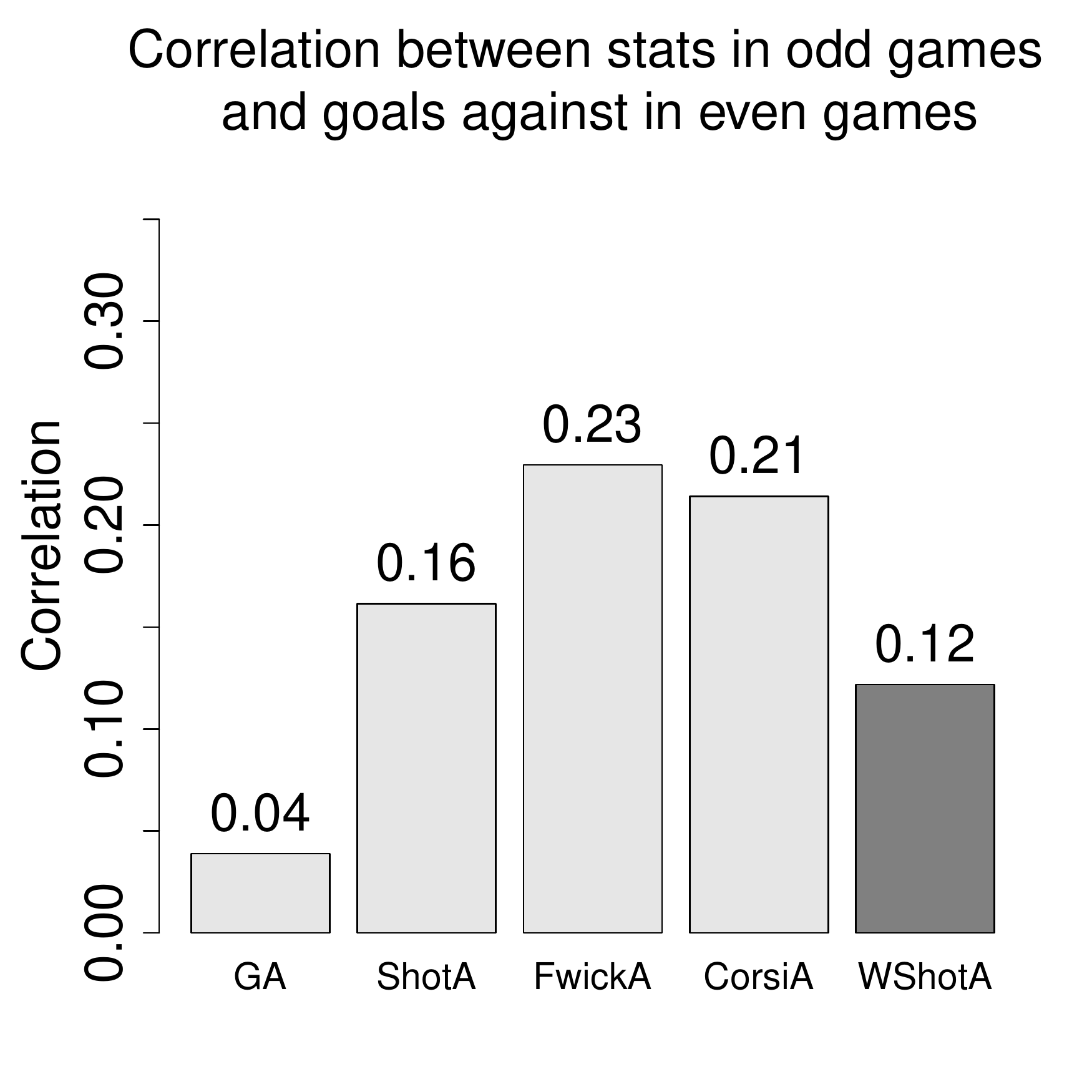}
\caption{(Left) Split-half reliability of various traditional and advanced per 60 minutes statistics at even strength.  (Right) Correlation between team statistics in odd games and goals against per 60 minutes in even games.  In both cases, only data from away games was used.}
        \label{team-GA60}
    \end{figure}

%

\newpage
\bibliographystyle{imsart-nameyear}
\bibliography{generalbib}

\begin{thebibliography}{28}

\bibitem[\protect\citeauthoryear{Awad}{2009}]{awadwshot}
\begin{bmisc}[author]
\bauthor{\bsnm{Awad},~\bfnm{Thomas}\binits{T.}}
(\byear{2009}).
\btitle{{Does Shot Quality Exist?}}
\bnote{\url{http://www.hockeyprospectus.com/article.php?articleid=540}.
  Accessed 12-22-2011.}
\end{bmisc}
\endbibitem

\bibitem[\protect\citeauthoryear{Awad}{2010}]{deltasot}
\begin{bmisc}[author]
\bauthor{\bsnm{Awad},~\bfnm{Tom}\binits{T.}}
(\byear{2010}).
\btitle{{Delta with Teammate Adjustments - DeltaSOT}}.
\bnote{\url{http://www.puckprospectus.com/article.php?articleid=454}, Accessed
  01-03-2012}.
\end{bmisc}
\endbibitem

\bibitem[\protect\citeauthoryear{Desjardins}{2010}]{gabe}
\begin{bmisc}[author]
\bauthor{\bsnm{Desjardins},~\bfnm{Gabriel}\binits{G.}}
(\byear{2010}).
\btitle{{Behind The Net}}.
\bnote{\url{http://www.behindthenet.ca}}.
\end{bmisc}
\endbibitem

\bibitem[\protect\citeauthoryear{Desjardins}{2011}]{gabe-shot-distance}
\begin{bmisc}[author]
\bauthor{\bsnm{Desjardins},~\bfnm{Gabriel}\binits{G.}}
(\byear{2011}).
\btitle{{Shot Distance Allowed as a Team Talent}}.
\bnote{\url{http://www.arcticicehockey.com/2011/10/24/2506209/shot-distance-allowed-as-a-team-talent}}.
\end{bmisc}
\endbibitem

\bibitem[\protect\citeauthoryear{Fearnhead and
  Taylor}{2011}]{fearnhead-taylor-nba}
\begin{barticle}[author]
\bauthor{\bsnm{Fearnhead},~\bfnm{Paul}\binits{P.}} \AND
  \bauthor{\bsnm{Taylor},~\bfnm{Benjamin~Matthew}\binits{B.~M.}}
(\byear{2011}).
\btitle{On Estimating the Ability of NBA Players}.
\bjournal{Journal of Quantitative Analysis in Sports}
\bvolume{7}
\bpages{11}.
Available at
  \url{http://EconPapers.repec.org/RePEc:bpj:jqsprt:v:7:y:2011:i:3:n:11}
\end{barticle}
\endbibitem

\bibitem[\protect\citeauthoryear{Fyffe and Vollman}{2002}]{fyffe-vollman}
\begin{bmisc}[author]
\bauthor{\bsnm{Fyffe},~\bfnm{Iain}\binits{I.}} \AND
  \bauthor{\bsnm{Vollman},~\bfnm{Rob}\binits{R.}}
(\byear{2002}).
\btitle{Improving Plus-Minus}.
\bnote{\url{http://hockeythink.com/research/plusmin.html}, Accessed
  01-03-2012}.
\end{bmisc}
\endbibitem

\bibitem[\protect\citeauthoryear{Hosmer and Lemeshow}{2000}]{hosmer-lemeshow}
\begin{bbook}[author]
\bauthor{\bsnm{Hosmer},~\bfnm{David~W.}\binits{D.~W.}} \AND
  \bauthor{\bsnm{Lemeshow},~\bfnm{Stanley}\binits{S.}}
(\byear{2000}).
\btitle{Applied logistic regression (Wiley Series in probability and
  statistics)}, \bedition{2} ed.
\bpublisher{Wiley-Interscience Publication}.
Available at
  \url{http://www.amazon.com/Applied-logistic-regression-probability-statistics/dp/0471356328%3FSubscriptionId%3D192BW6DQ43CK9FN0ZGG2%26tag%3Dws%26linkCode%3Dxm2%26camp%3D2025%26creative%3D165953%26creativeASIN%3D0471356328}
\end{bbook}
\endbibitem

\bibitem[\protect\citeauthoryear{Ilardi and Barzilai}{2008}]{ilardibarzilai}
\begin{bmisc}[author]
\bauthor{\bsnm{Ilardi},~\bfnm{S.}\binits{S.}} \AND
  \bauthor{\bsnm{Barzilai},~\bfnm{A.}\binits{A.}}
(\byear{2008}).
\btitle{{Adjusted Plus-Minus Ratings: New and Improved for 2007-2008}}.
\bnote{\url{http://www.82games.com/ilardi2.htm}}.
\end{bmisc}
\endbibitem

\bibitem[\protect\citeauthoryear{Johns}{2004}]{johns}
\begin{bmisc}[author]
\bauthor{\bsnm{Johns},~\bfnm{Graeme}\binits{G.}}
(\byear{2004}).
\btitle{{Statistical Shot Quality Weighting}}.
\bnote{\url{http://hockeyanalytics.com/2004/10/statistical-shot-quality-weighing/}}.
\end{bmisc}
\endbibitem

\bibitem[\protect\citeauthoryear{Johnson}{}]{davidjohnson}
\begin{bmisc}[author]
\bauthor{\bsnm{Johnson},~\bfnm{David}\binits{D.}}
\btitle{Hockey Analysis Player Ratings}.
\bnote{\url{http://stats.hockeyanalysis.com/about.php}, Accessed 01-03-2012}.
\end{bmisc}
\endbibitem

\bibitem[\protect\citeauthoryear{Krzywicki}{2005}]{ken1}
\begin{bmisc}[author]
\bauthor{\bsnm{Krzywicki},~\bfnm{Ken}\binits{K.}}
(\byear{2005}).
\btitle{{Shot Quality Model: A logistic regression approach to assessing NHL
  shots on goal}}.
\bnote{\url{http://www.hockeyanalytics.com/Research_files/Shot_Quality_Krzywicki.pdf}}.
\end{bmisc}
\endbibitem

\bibitem[\protect\citeauthoryear{Krzywicki}{2009}]{ken2}
\begin{bmisc}[author]
\bauthor{\bsnm{Krzywicki},~\bfnm{Ken}\binits{K.}}
(\byear{2009}).
\btitle{{Removing Observer Bias from Shot Distance - Shot Quality Model - {NHL}
  Regular Season 2008-09}}.
\bnote{\url{http://www.hockeyanalytics.com/Research_files/SQ-DistAdj-RS0809-Krzywicki.pdf}}.
\end{bmisc}
\endbibitem

\bibitem[\protect\citeauthoryear{Krzywicki}{2010}]{ken3}
\begin{bmisc}[author]
\bauthor{\bsnm{Krzywicki},~\bfnm{Ken}\binits{K.}}
(\byear{2010}).
\btitle{{NHL Shot Quality 2009-10: A look at shot angles and rebounds}}.
\bnote{\url{http://hockeyanalytics.com/2010/10/nhl-shot-quality-2010/}}.
\end{bmisc}
\endbibitem

\bibitem[\protect\citeauthoryear{Lewin}{2007}]{lewin}
\begin{bmisc}[author]
\bauthor{\bsnm{Lewin},~\bfnm{David}\binits{D.}}
(\byear{2007}).
\btitle{{2004-2005 Adjusted Plus-Minus Ratings}}.
\bnote{\url{http://www.82games.com/lewin3.htm}}.
\end{bmisc}
\endbibitem

\bibitem[\protect\citeauthoryear{Macdonald}{2011a}]{apm}
\begin{barticle}[author]
\bauthor{\bsnm{Macdonald},~\bfnm{Brian}\binits{B.}}
(\byear{2011}a).
\btitle{{A Regression-Based Adjusted Plus-Minus Statistic for NHL Players}}.
\bjournal{Journal of Quantitative Analysis in Sports}
\bvolume{7}
\bpages{29}.
Available at \url{www.bepress.com/jqas/vol7/iss3/4/}
\end{barticle}
\endbibitem

\bibitem[\protect\citeauthoryear{Macdonald}{2011b}]{ridge}
\begin{bmisc}[author]
\bauthor{\bsnm{Macdonald},~\bfnm{Brian}\binits{B.}}
(\byear{2011}b).
\btitle{{Adjusted Plus-Minus for NHL Players using Ridge Regression}}.
\bnote{{Submitted. ArXiv preprint: \url{http://arxiv.org/abs/1201.0317}}}.
\end{bmisc}
\endbibitem

\bibitem[\protect\citeauthoryear{Macdonald}{2011c}]{apm2}
\begin{barticle}[author]
\bauthor{\bsnm{Macdonald},~\bfnm{Brian}\binits{B.}}
(\byear{2011}c).
\btitle{{An Improved Adjusted Plus-Minus Statistic for NHL Players}}.
\bjournal{Proceedings of the MIT Sloan Sports Analytics Conference}.
Available at \url{http://www.sloansportsconference.com/?p=2838}
\end{barticle}
\endbibitem

\bibitem[\protect\citeauthoryear{Macdonald}{2012}]{spm}
\begin{barticle}[author]
\bauthor{\bsnm{Macdonald},~\bfnm{Brian}\binits{B.}}
(\byear{2012}).
\btitle{{An Expected Goals Model for Evaluating NHL Teams and Players}}.
\bjournal{Proceedings of the 2012 MIT Sloan Sports Analytics Conference}.
\bnote{\url{http://www.sloansportsconference.com/?p=6157}, Accessed 2-20-2012}.
\end{barticle}
\endbibitem

\bibitem[\protect\citeauthoryear{Macdonald, Arney and Peterson}{2012}]{coop}
\begin{barticle}[author]
\bauthor{\bsnm{Macdonald},~\bfnm{Brian}\binits{B.}},
  \bauthor{\bsnm{Arney},~\bfnm{Chris}\binits{C.}} \AND
  \bauthor{\bsnm{Peterson},~\bfnm{Elisha}\binits{E.}}
(\byear{2012}).
\btitle{{Modeling Cooperation in Networks, Organizations, and Systems}}.
\bjournal{in preparation}.
\end{barticle}
\endbibitem

\bibitem[\protect\citeauthoryear{Rosenbaum}{2004}]{rosenbaum}
\begin{bmisc}[author]
\bauthor{\bsnm{Rosenbaum},~\bfnm{Dan}\binits{D.}}
(\byear{2004}).
\btitle{{Measuring How NBA Players Help Their Teams Win}}.
\bnote{\url{http://www.82games.com/comm30.htm}}.
\end{bmisc}
\endbibitem

\bibitem[\protect\citeauthoryear{Ryder}{2004}]{ryder-shot-quality}
\begin{bmisc}[author]
\bauthor{\bsnm{Ryder},~\bfnm{Alan}\binits{A.}}
(\byear{2004}).
\btitle{{Isolating Shot Quality}}.
\bnote{\url{http://hockeyanalytics.com/2004/01/isolating-shot-quality/}}.
\end{bmisc}
\endbibitem

\bibitem[\protect\citeauthoryear{Ryder}{2007}]{ryder-shot-quality-recall}
\begin{bmisc}[author]
\bauthor{\bsnm{Ryder},~\bfnm{Alan}\binits{A.}}
(\byear{2007}).
\btitle{{Product Recall Notice for Shot Quality}}.
\bnote{\url{http://hockeyanalytics.com/2007/06/product-recall-notice-for-shot-quality/}}.
\end{bmisc}
\endbibitem

\bibitem[\protect\citeauthoryear{Schuckers}{2011}]{digr}
\begin{bmisc}[author]
\bauthor{\bsnm{Schuckers},~\bfnm{Michael}\binits{M.}}
(\byear{2011}).
\btitle{{DIGR: A Defense Independent Rating of NHL Goaltenders using Spatially
  Smoothed Save Percentage Maps}}.
\bnote{\url{http://www.sloansportsconference.com/?p=648}}.
\end{bmisc}
\endbibitem

\bibitem[\protect\citeauthoryear{Seppa}{2009}]{seppa}
\begin{bmisc}[author]
\bauthor{\bsnm{Seppa},~\bfnm{Timo}\binits{T.}}
(\byear{2009}).
\btitle{{Even Strength Total Rating}}.
\bnote{\url{http://www.puckprospectus.com/article.php?articleid=254}}.
\end{bmisc}
\endbibitem

\bibitem[\protect\citeauthoryear{Sill}{2010}]{sill}
\begin{barticle}[author]
\bauthor{\bsnm{Sill},~\bfnm{Joe}\binits{J.}}
(\byear{2010}).
\btitle{{Improved NBA Adjusted +/- Using Regularization and Out-of-Sample
  Testing}}.
\bjournal{Proceedings of the 2010 MIT Sloan Sports Analytics Conference}.
\end{barticle}
\endbibitem

\bibitem[\protect\citeauthoryear{Tango}{2010}]{tango-wowy}
\begin{bmisc}[author]
\bauthor{\bsnm{Tango},~\bfnm{Tom}\binits{T.}}
(\byear{2010}).
\btitle{With or Without You - At the win/loss game level}.
\bnote{\url{http://www.insidethebook.com/ee/index.php/site/article/with_or_without_you_at_the_win_loss_game_level/},
  Accessed 01-03-2012}.
\end{bmisc}
\endbibitem

\bibitem[\protect\citeauthoryear{Wilson}{2011}]{wilson-wowy}
\begin{bmisc}[author]
\bauthor{\bsnm{Wilson},~\bfnm{Kent}\binits{K.}}
(\byear{2011}).
\btitle{{With or Without Bobby Ryan}}.
\bnote{\url{http://www.hockeyprospectus.com/article.php?articleid=763},
  Accessed 01-03-2012}.
\end{bmisc}
\endbibitem

\bibitem[\protect\citeauthoryear{Witus}{2008}]{eli}
\begin{bmisc}[author]
\bauthor{\bsnm{Witus},~\bfnm{Eli}\binits{E.}}
(\byear{2008}).
\btitle{{Count the Basket}}.
\bnote{\url{http://www.countthebasket.com/blog/}}.
\end{bmisc}
\endbibitem

\end{thebibliography}
%
%
%
%
%
%

\end{document}